\documentclass[aps,pre,twocolumn,graphicx]{revtex4-1}

\usepackage[utf8]{inputenc}
\usepackage[T1]{fontenc}
\usepackage{amsmath}  % needed for \tfrac, \bmatrix, etc.
\usepackage{amsfonts} % needed for bold Greek, Fraktur, and blackboard bold
\usepackage{graphicx} % needed for figures
\usepackage{siunitx}
\usepackage{color}

\usepackage{subcaption}

\usepackage{hyperref}
\usepackage{url}
\usepackage{cleveref}

\graphicspath{{figures/}}

\begin{document}

\title{Shape, connectedness percolation and electrical conductivity of clusters in suspensions of hard platelets}

\author{Arshia Atashpendar}
\email{arshia.atashpendar@physik.uni-freiburg.de}
\affiliation{Institute of Physics, University of Freiburg, Hermann-Herder-Str.\ 3, 79104 Freiburg, Germany}

\author{Tim Ingenbrand}
%\email{}
\affiliation{Institute of Physics, University of Freiburg, Hermann-Herder-Str.\ 3, 79104 Freiburg, Germany}

\author{Tanja Schilling}
\affiliation{Institute of Physics, University of Freiburg, Hermann-Herder-Str.\ 3, 79104 Freiburg, Germany}

\date{\today}
\begin{abstract}

Using Monte Carlo simulations, we investigate how geometric percolation and electrical conductivity in suspensions of hard conducting platelets are affected by the addition of platelets and their degree of spontaneous alignment. In our simulation results for aspect ratios $10, 25$ and $50$, we consistently observe a monotonically decreasing percolation threshold as a function of volume fraction, i.e., the addition of particles always aids percolation. In the nematic phase, the distribution of particles inside the percolating clusters becomes less spherically symmetric and the aspect ratio of the clusters increases. However, the clusters are also anisotropically shaped in the isotropic phase, although their aspect ratio remains constant as a function of volume fraction and is only weakly dependent on the particle aspect ratio. Mapping the percolating clusters of platelets to linear resistor networks, and assigning unit conductance to all connections, we find a constant conductivity both across the isotropic-nematic transition and in the respective stable phases. This behaviour is consistent with the other observed topological properties of the networks, namely, the average path length, average number of contacts per particle and the Kirchhoff index which all remain constant and unaffected by both the addition of particles and the degree of alignment of their suspension. On the contrary, using an anisotropic conductance model that explicitly accounts for the relative orientation of the particles, the network conductivity decreases with increasing volume fraction in the isotropic, and further diminishes at the onset of the nematic while preserving the same trend deep in the nematic.  Hence, our observations consistently suggest that unlike for rod-like fillers, the network structures that arise from platelet suspensions are neither very sensitive to the particle aspect ratio nor to alignment. Hence platelets are not as versatile as fillers for dispersion in conductive composite materials as rods.

\end{abstract}

%\pacs{72.80.Tm, 64.70.M-, 73.40.Gk, 64.60.ah}% insert suggested PACS numbers in braces on next line

\maketitle %\maketitle must follow title, authors, abstract and \pacs

\section{Introduction}

The dispersion of composite polymeric systems with nano-particles, which are electrically connected by quantum tunneling of electrons, has become a useful route to the production of conductive plastic materials. Commonly, anisotropic fillers particles are of interest as the load required to achieve a percolating network can be lowered with increasing aspect ratio of the particles. This behaviour is known to be exhibited by rods~\cite{balberg1984, MUTISO201563, Schilling_2015, Ambrosetti2010perc}, and by platelets such as oblate ellipsoids~\cite{ambrosetti2008oblates}, unless the percolation transition is preempted by a spontaneous formation of a nematic liquid crystal phase~\cite{MathewPercolation}, thereby, the monotonic decrease of the required load of platelets as a function of aspect ratio may be lost.

For platelets, dispersions of typically graphene but also clay are used for enhancing the electrical conductivity of nanocomposites~\cite{POTTS20115, FELLER2004739, OKAMOTO20012685, yousefi2012self, doi:10.1021nn500590g, doi:10.1021nn303904z}, next to various other technological applications such as corrosion protection~\cite{Clay2003corrosion}, and electromagnetic interference shielding~\cite{LEE201757}. Moreover, there also exists various types of dispersions, e.g., using functionalized filler platelets to control their emergent network morphology~\cite{doi:10.1021/la0349267}, polydisperse suspensions exhibiting quasi-universal dependence on the filler aspect ratio~\cite{kale2015polydisperse}, or mixtures of different types of fillers to exploit their synergistic effects~\cite{kale2018multicomp}.

Another important aspect of anisotropic fillers such as rods or platelets is their ability to intrinsically form liquid crystalline phases: upon increasing the concentration of the particles, they undergo a first order transition from an isotropic (I) to a nematic (N) phase~\cite{onsager1949effects, p1995physics} where the particles become \textit{orientationally} ordered. For rod-like particles, the effect of the so-induced alignment on the tunneling conductivity has been previously studied in simulations~\cite{ATunnelingRods}, using theoretical percolation-based approaches in~\cite{Chatterjee_2015, PhysRevE.98.062102} for various degrees of orientational order, and more extensively in cases of externally induced alignment in~\cite{ackermann2016effect, wang2008effects, finner2018continuum, kale2016effect, lebovka2018anisotropy}. However for platelets, the percolation and conductivity behaviour across the intrinsic liquid-crystalline phases has remained largely unexplored. Here, we aim to tackle this topic using computer simulations. 

For the dispersion of conductive nanofillers to globally enhance the conductivity of the host system, the desired dispersion must be electrically percolating. This means, conductive pathways formed by the particle clusters can be found between almost any two junctions of the host composite system. Therefore, before tackling the conductivity aspects of such dispersions, it is prerequisite to understand how system-spanning clusters of platelets are formed, and how their shape is affected by the degree of alignment of the suspension under thermodynamic equilibrium conditions. 
In the case of electrical connectivity established by tunneling, a given pair of particles are said to be \textit{connected} if their shortest surface to surface distance is on the same order of magnitude as the characteristic tunneling decay length of particles in the host medium. The latter is typically modeled by defining a geometric connectivity criterion~\cite{Balberg_2009, doi:10.1063/1.3559004, Kyrylyuk8221} where the particles are assumed to be coated by penetrable shells, and thus, \textit{to be electrically connected} translates into having adjacent particles with overlapping shells. The connection to tunneling is completed by interpreting the particle shell thickness as their tunneling decay length~\cite{Ambrosetti2010perc, Balberg_2009, PhysRevB.74.054205}.

Adopting the latter connectivity criterion, our first aim in this article is to study the geometric percolation of suspensions of hard platelets across their intrinsic I-N transition. More precisely, this entails estimating the percolation threshold (i.e., critical shell size) as a function of volume fraction and quantifying the shape of the percolating clusters. 

Subsequently, viewing the so-defined particle connections as resistors, we map the tunneling-based networks of platelets to linear resistor networks and study their conductivity along with other topological properties such as the Kirchhoff index and the average path length. In order to perform this mapping, we use various conductance (resistor) definitions: a simple network where a unit conductance is assigned to all particle connections, and an anisotropic model where the conductance is weighed according to the relative orientation of the platelets and their respective center of mass positions. The latter model is defined such that maximum conductance is assigned to connected pairs that are both parallel and have their centers aligned, while perpendicularly oriented and end-to-end connected pairs are assigned intermediate and minimum values respectively. The unit conductance model serves to establish a base-line in the conductivity behaviour in terms of the average number of contacts per particle, the typical number of hops between two arbitrary particles in a cluster and the respective number of pathways connecting them, all of which may be affected by the liquid-crystalline phase of the suspension. In contrast, the motivation for the anisotropic conductance model lies in the tunneling anisotropy of anisotropic filler particles, which will be further detailed in the subsequent section.

The remainder of this article is structured as follows: In Sec.~\ref{sec:models}, we further elaborate on the models, detail our Monte Carlo (MC) simulations, and explain our computational methods. In Sec.~\ref{sec:results}, we present our results by starting from the shape quantification of the percolating clusters and the corresponding critical shell sizes as function of volume fraction. Then, we discuss our findings on the electrical conductivity and topology of the tunneling networks. Finally, Sec.~\ref{sec:summary} summarizes our main results along with discussions on a potential line of continuation for future work.

\section{Models and Simulations}\label{sec:models}

We used cut-spheres as a model for disk-like particles. A cut-sphere is characterized by its diameter $D$  and thickness $L$ and it can be obtained by starting from a sphere of diameter $D$ and removing those parts of the sphere that are a distance $L/2$ above and below the equatorial plane. From the computational aspect of modeling platelets, cut-spheres are advantageous over for instance oblate spherocylinders or cylinders, as testing for overlaps between a given pair can be carried out in a finite number of steps~\cite{allen1993hardconvex}. The latter is particularly important when hard interactions are of interest, meaning overlapping configurations of platelets are forbidden (interaction potential $\to \infty$) and non-overlapping ones are allowed with probability $1$ (interaction potential $=0$). Throughout this article, we refer to cut-spheres as platelets. 

The hard core of a platelet is centered inside an imaginary shell of the same shape. The imaginary coating serves to define a geometric connectivity criterion: Given a shell size $A,$ two platelets are considered to be connected if their shells overlap, i.e. their surface-to-surface distance is smaller than $A.$ A contiguous sequence of so-connected platelets then form a cluster. Percolation occurs when a cluster is wrapping the simulation box~\cite{vskvor2007} through at least one direction of the periodic boundaries. The so-defined geometric percolation can also be related to electrical percolation when the shell thickness $A$ is interpreted as the tunneling distance, i.e., the distance at which the electron tunneling probability is reduced to $1/e$ times its initial value.

Using canonical (NVT) Monte Carlo (MC) simulations, we have generated equilibrated configurations of hard platelets. The simulations have been carried out for monodisperse systems of aspect ratio $D/L=10, 25$ and $50,$ using a cubic simulation box of dimensions $\mathcal{L}_{x,y,z}=16 D,$ with periodic boundary conditions in all $3$ dimensions. The particle numbers were chosen in a range such that the volume fraction spans the stable regions of both the isotropic and nematic phases of the suspensions of hard platelets. The volume fraction $\phi$ is defined as $\phi=Nv_{\rm core}/V,$ with $N$ the number of particles, $V$ the volume of the simulation box and $v_{\rm core}$ denotes the volume of one platelet given by

\begin{equation}\label{eq:volplatelet}
v_{\rm core} = \frac{\pi L}{4}\left(D^2 - \frac{L^2}{3} \right).
\end{equation}

In the infinite system size limit, the percolation wrapping probability behaves as a Heaviside step function with a jump at the critical volume fraction $\phi_p$ or shell size $A_p.$ However for finite systems, the transition is often described as being ``smeared out'', where briefly, the asymptotic finite size scaling of a system observable $O$ is given by~\cite{aharony:2003}

\begin{equation}\label{eq:fss}
    O(\mathcal{L},\lambda) \propto F(\delta \lambda \mathcal{L}^{1/\nu})
\end{equation}
with $\delta \lambda = \lambda - \lambda_p$ the distance of the scaling parameter (e.g., the shell size $A$) from its critical value, $\mathcal{L}$ the linear size of the system (i.e., simulation box length), $F$ the appropriate scaling function for the chosen observable (here wrapping probability) and $\nu$ the critical exponent of the characteristic correlation length $\xi \propto |\delta \lambda|^{-\nu}.$ With the hypothesis of a one parameter scaling law as given by Eq.~\ref{eq:fss}, it is clear that by studying the wrapping probability as a function of finite system sizes $\mathcal{L},$ all the obtained curves cross at a common intersection point $F(\delta \lambda=0),$ which allows us to estimate the critical value corresponding to the limit of $\mathcal{L}\to \infty$ without requiring any assumptions on the critical exponents. 

Typically, the finite size scaling analysis then entails simulating a range of different simulation box sizes, performing a fine sweep in the chosen control parameter of percolation and determining the critical value near the transition by estimating the common crossing point of their respective $p$-curves. 
However, for the used box sizes (e.g., $\mathcal{L}=16D$), our observed finite size effects remain negligibly small in both the isotropic and nematic phases, and a common intersection point can always be found without a systematic shift of the curves as a function of system size. Thus, to reduce computational costs, we decided to perform the sweep in shell size $A$ only for one system size per $\phi$ and use the point $p(A_p) \approx 0.5$ as an estimate of the percolation threshold $A_p.$ The same approach is used in order to estimate the trend of $A_p$ as a function of $\phi,$ as the impact of finite-size effects on the percolation probability remains negligible compared to that of $A$ or $\phi.$

The nematic order parameter $S_2$ is used in order to distinguish between the isotropic and nematic phase of the platelets, given by the largest eigenvalue of the orientation tensor $Q:$

\begin{equation}\label{eq:s2}
Q_{ij} = \frac{1}{2N} \sum_{\alpha=1}^N (3v_i^\alpha v_j^\alpha-\delta_{ij}),
\end{equation}

where $v_i^\alpha$ and $v_j^\alpha$ are the ith and jth components of the normalized orientation vector of platelet $\alpha,$ respectively, $N$ the particle number, and $\delta_{ij}$ is the Kronecker delta. Additionally, the stable phase boundaries are determined according to the following criteria of the $S_2$ values upon equilibration: suspensions with $|S_2| <0.05$ lie in the stable isotropic region, while $S_2>0.5$ correspond to the stable nematic, and values in between are deemed metastable, thus, roughly indicating the coexistence window.

\begin{figure}
    \centering
      \begin{subfigure}{0.2\textwidth}
        \includegraphics[width=0.75\textwidth]{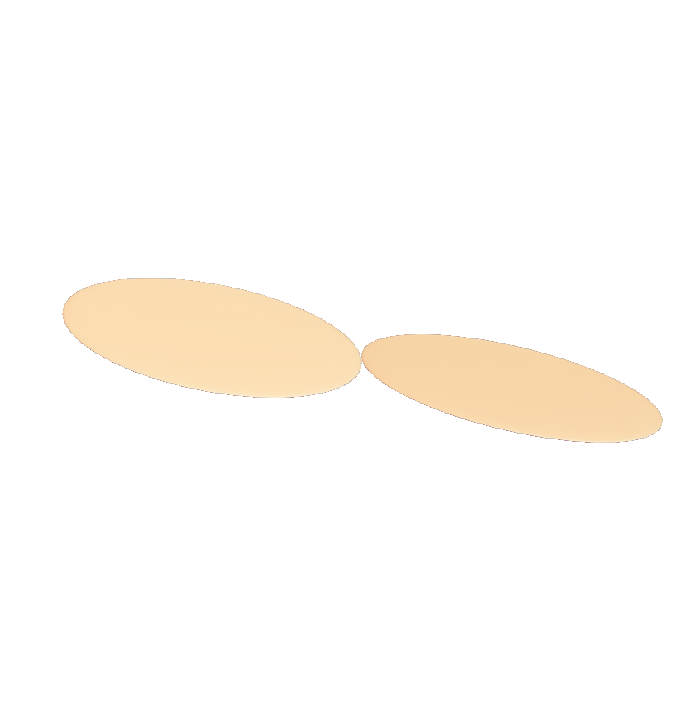}
          \caption{}
          \label{fig:snapa}
      \end{subfigure}
      \hfill
      \begin{subfigure}{0.2\textwidth}
        \includegraphics[width=0.75\textwidth]{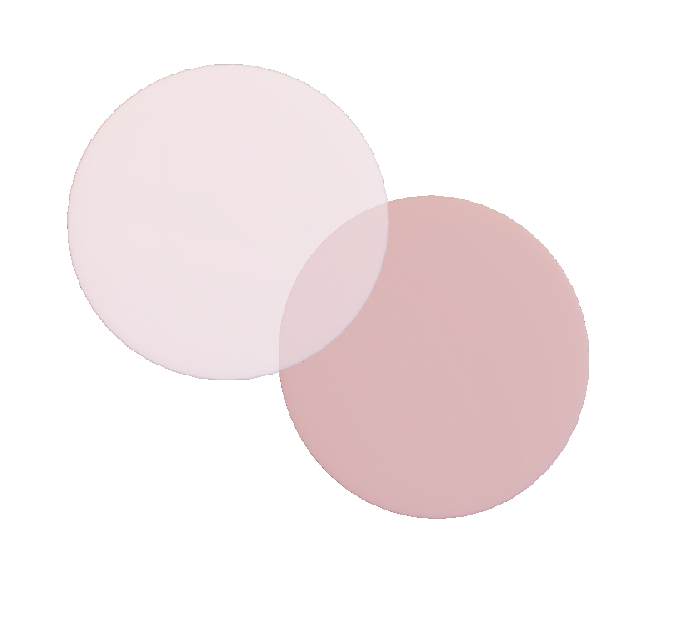}
         \caption{}
          \label{fig:snapb}
      \end{subfigure}
      \begin{subfigure}{0.2\textwidth}
        \includegraphics[width=0.75\textwidth]{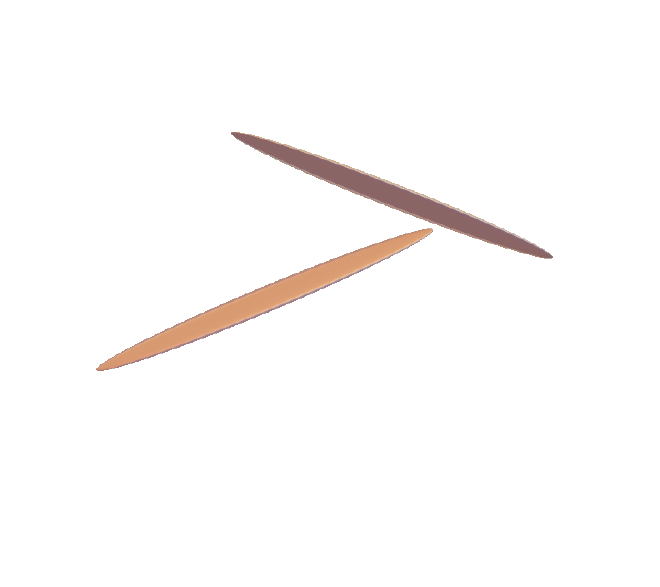}
          \caption{}
          \label{fig:snapc}
      \end{subfigure}
      \begin{subfigure}{0.2\textwidth}
        \includegraphics[width=0.75\textwidth]{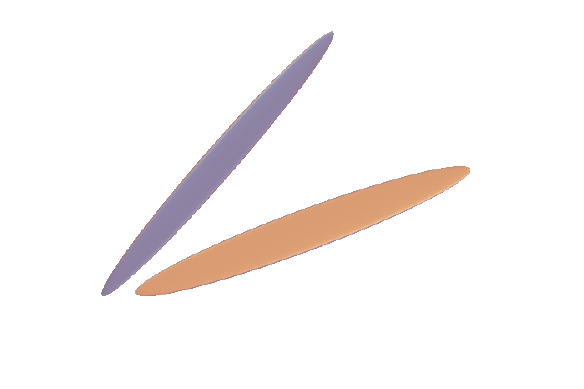}
          \caption{}
          \label{fig:snapd}
      \end{subfigure}
      \begin{subfigure}{0.2\textwidth}
        \includegraphics[width=0.75\textwidth]{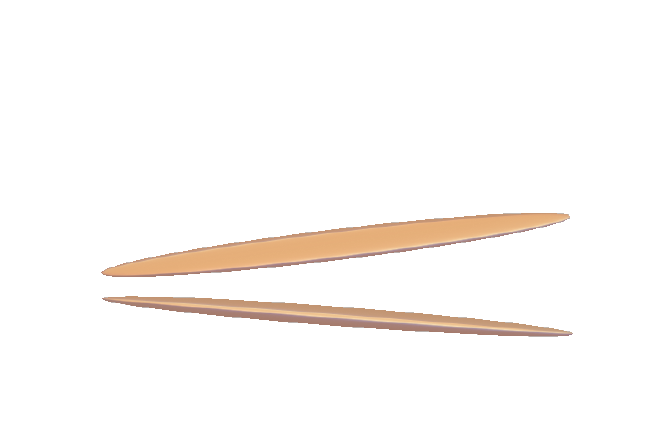}
         \caption{}
          \label{fig:snape}
      \end{subfigure}
\caption{
\label{fig:snapall}%
Snapshots of typical configurations of connected pairs of platelets. 
For comparison, the corresponding conductance weights computed according to the simple geometric model $w_a$ and the normalized cross section $w_s$ are: $(a): w_a\approx 0.067, w_s\approx 0.0012,$ $(b): w_a\approx 0.12, w_s\approx 0.15,$ $(c): w_a\approx 0.46, w_s\approx 0.41,$ $(d): w_a\approx 0.89, w_s\approx 0.78,$ and $(e): w_a\approx 0.99, w_s\approx 0.96.$
}
\end{figure}

\begin{figure}
\centering
\includegraphics[width=0.49\linewidth]{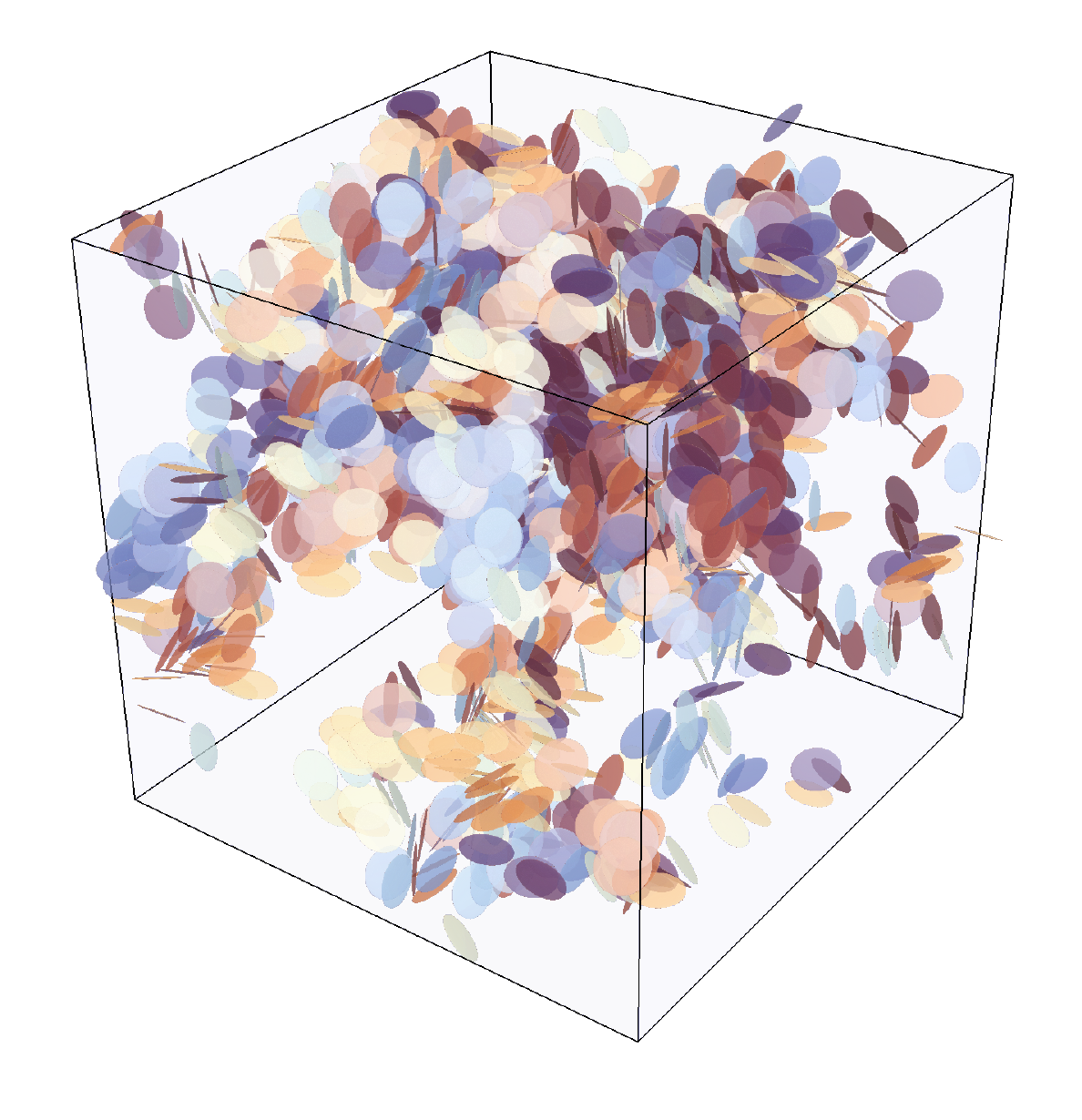}
\includegraphics[width=0.49\linewidth]{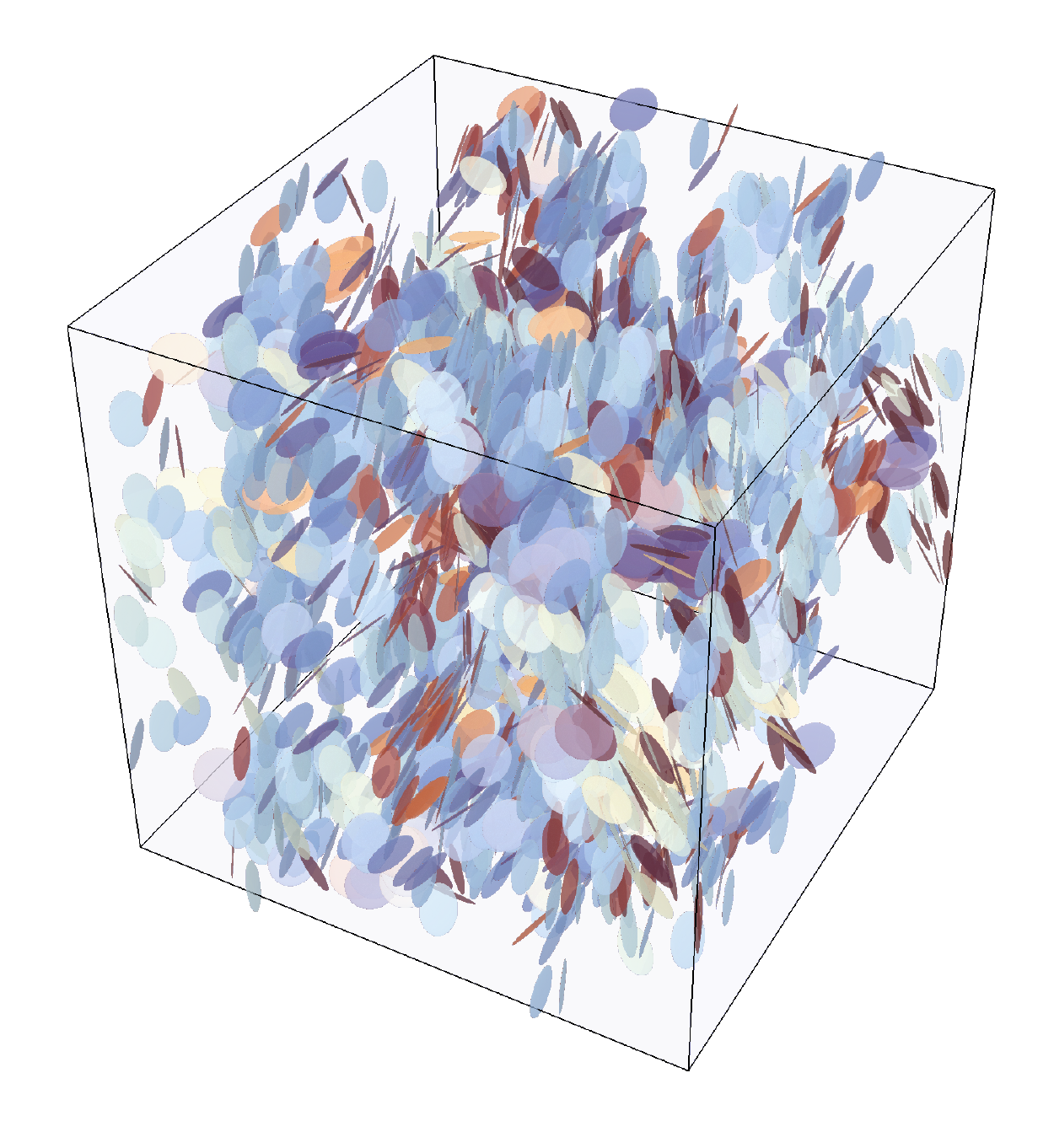}
\caption{Largest clusters of platelets (drawn as oblates) in isotropic (left) and nematic (right) suspensions visualised for $D/L=25.$}
\label{fig:DL25isonemclus}
\end{figure}

Next, we briefly describe how our conductivity calculations are performed. Given an equilibrated and percolating suspension, we extract its largest cluster (examples are visualised in Fig.~\ref{fig:DL25isonemclus}) and view its underlying connectivity network as an undirected, simple and connected graph $G=(V,E).$ The set of vertices $V$ is comprised of a vertex assigned to each platelet in the cluster, and the edge set $E$ is defined according to the connectivity criterion of overlapping shells, i.e., $e_{ij}\in E$ for each pair of platelets $i$ and $j$ with overlapping shells. The mapping to a resistor network is completed by considering the triple $(V,E,w),$ where $w$ is a weight function that assigns a positive conductance (or resistance $r=1/w$) to every edge $e_{ij}\in E.$ Furthermore, we only consider the case of \textit{linear} resistor networks, meaning each edge is a linear resistor where the current through it and the voltage across the edge are related by Ohm's law, $I_{i,j}=V_{i,j}/r_{i,j},$ with $r_{i,j}>0$ the resistance of the edge between the $i$ and $j$ junctions.  For defining the conductance $w_{ij}$ between a given pair of connected platelets, we use the following two models:

\begin{align}
w_{u}(e_{ij}) :=& 1, \text{ for all }e_{ij}\in E, \\
w_{a}(e_{ij}) :=& \sqrt{|\mathbf{m_i} \cdot \hat{\mathbf{r}} * \mathbf{m_j} \cdot \hat{\mathbf{r}}|,} \label{eq:geomstack}
\end{align}

where $\hat{\mathbf{r}}$ is the normalized difference vector between the center of mass vectors $\mathbf{r_{i,j}}$ of the platelets, and $\mathbf{m_{i,j}}$ are their respective orientations: unit vectors pointing along the short axis of the platelets. The specific functional form of the product of relative angles in Eq.~\ref{eq:geomstack} is chosen such that it roughly satisfies $w_a \approx 1$ for platelets that are aligned and centered (example Fig.~\ref{fig:snape}), $w_a \approx 0.5 $ for one nearly perpendicular to the other (Fig.~\ref{fig:snapc}) and $w_a\approx 0.1-0.3$ for cases of minimal overlap, such as side-to-side connected and aligned platelets (Fig.~\ref{fig:snapa}, \ref{fig:snapb}). 

The so-defined $w_a$ represents a geometric and easy-to-compute alternative to a weight function that explicitly involves calculating the cross section area $w_s$ of two platelets. For a selection of cases, the latter has been computed in order to ensure Eq.~\ref{eq:geomstack} correctly approximates $w_s$ in assigning conductances based on the described ways a pair can be connected. Comparisons are shown in Fig.~\ref{fig:snapall}, with $w_s$ normalized by $\pi D^2/4.$

The cross section area $w_s$ between two platelets is computed by taking the mean overlapping areas obtained from the projection of the disk-shaped surface of one platelet onto the plane of the other and vice versa. The respective intersecting area upon each projection is computed numerically. Unfortunately, performing the entirety of our conductivity calculations with the latter approach would be computationally infeasible, therefore, our main measurements are conducted either by uniformly assigning unit conductances or using Eq.~\ref{eq:geomstack}.

The motivation underlying the non-uniform conductance models stems from the orientational dependence of quantum tunneling of electrons between anisotropic filler particles. Particularly for rod-like filler particles, G. Nigro and C. Grimaldi~\cite{NigroGrimaldi} have shown that in general the tunneling between parallel rods is ($L/\sqrt{D\xi}$\footnote{where $D$ and $L>>D$ are the diameter and length of the rodlike particle respectively, and $\xi$ denotes the tunneling decay length.} times) larger than the tunneling between perpendicular ones. Their result follows from the consideration that overlap between the wavefunctions of two rods is largest when the overlap extends over the entire long-axis length of the rods. 

As far as we know, a similar and complete quantum mechanical treatment of tunneling between platelet-like particles has not yet been performed. Nonetheless, in direct analogy with the solved case of rods~\cite{NigroGrimaldi}, it stands to reason that two platelets whose centers are aligned and have parallel orientations, exhibit maximally overlapping wavefunctions as their cross section extends over the entire surface of the platelets, i.e., the most favourable setup for tunneling. Similarly, minimal overlap would follow for their end-to-end connection and intermediate to these two cases would for example be side-to-end connections (perpendicular orientation vectors).

With the above setting for defining electrical networks from clusters of platelets, our conductivity calculations amount to computing the effective resistance, also simply called resistance distance~\cite{Klein1993}, between arbitrary junctions of the network. Thus, we recur to the study of resistive electrical networks in algebraic graph theory and circuit theory. Below we will provide a brief description for the computation of the effective resistance and current flow across the network, but for a comprehensive overview we refer the reader to the following pieces of literature~\cite{Bollobas1998, Redner2009, ellens2011effective}, and in particular~\cite{DorflerElectrical} for the linear algebraic formulation of Kirchhoff laws for graphs.

It is useful to assign arbitrary orientations to each edge $e\in E,$ which set the reference direction of current flow through an edge. Then, the incidence matrix of the resulting directed graph operates as a difference matrix which allows for a simple formulation of Kirchhoff's current and voltage laws. As a convention, positive and negative currents represent a flow entering and exiting a node, respectively. The so defined directed graph, with $n$ nodes and $m$ edges, is characterized by an $n$ by $m$ incidence matrix $B,$ with each column of $B$ corresponding to an edge $(i,j),$ which contains a $1$ at the $i$th-row and $-1$ at the $j$th-row. With the incidence matrix $B$ and the $m$ by $m$ diagonal matrix of conductances $W,$ the Kirchhoff matrix or the Laplacian matrix of the graph $G$ is given by 

\begin{equation}\label{eq:laplacian}
K = BWB^T.
\end{equation} 

With the Laplacian, the Kirchhoff equations relating the \textit{nodal} current and voltage vectors $\mathbf{J_n}$ and $\mathbf{V_n},$ is given by $\mathbf{J_n} = K\mathbf{V_n},$ while Ohm's law relates the vector of currents $\mathbf{J_e}$ through the edges with the vector of voltage differences $\mathbf{V_e}$ across the edges $\mathbf{J_e}=W \mathbf{V_e}.$ In order to calculate the resistance distance $R_{ab}$ between arbitrary nodes $a,b \in V(G)$ we first supply the network with a current $\mathbf{J_n}=\mathbf{q}$ and measure the voltage difference across the nodes $a$ and $b.$ More precisely, an external unit current $i_{\mathrm{ext}}=1$ is inserted at the node $a,$ and extracted at the node $b,$ i.e., $\mathbf{J_n}^a=i_{\mathrm{ext}}$ and $\mathbf{J_n}^b=-i_{\mathrm{ext}},$ and zero for the remaining nodes $c\in V(G)\setminus \{a,b\}$ not connected to a current supply, i.e., $\mathbf{J_n}^c=\mathbf{q}^c=0.$ Then, $R_{ab}$ simply corresponds to the ratio of the potential difference across the two nodes with the net current: $R_{ab} = \frac{\mathbf{V_n}^a - \mathbf{V_n}^b}{i_{\mathrm{ext}}}.$ The potentials at the nodes are found by solving the linear system $K \mathbf{V_n}=\mathbf{J_n}=\mathbf{q},$ for which we recur to the method of direct linear solvers for Laplacian systems~\cite{VishnoiLaplacian}. 

Considering the fact that the Laplacian $K$ is a \textit{singular} matrix (since $\operatorname{Ker}(K)=\mathbf{1}$), the linear system is not directly invertible and can only be solved when $\mathbf{q}\in \operatorname{Im}(K),$ and by computing the pseudo-inverse $K^+$ of the Laplacian.  The latter is achieved by first casting the Laplacian system to an equivalent saddle point problem~\cite{schonknecht2015solvers, benzi_golub_liesen_2005}, and the matrix inversion obtained using UMFPACK~\cite{Davis:2004:AUV:992200.992206} as a sparse direct solver.   

Alternatively, and typically for larger networks, one may also apply a Star-Mesh transform (see e.g., the Appendix in~\cite{NigroGrimaldiDepletion}) to the network until only the electrode nodes remain, yielding the effective resistance between them. This is a special case of the Kron reduction~\cite{DorflerKron}, which entails finding a reduced equivalent network by taking the Schur complement of the Laplacian (which is an M-matrix) with respect to a subset of boundary nodes (e.g. two electrode nodes). Finally, in order to study the flow of current across the network, we compute the current through each edge according to $\mathbf{J_e}=W\mathbf{V_e} = WB^T \mathbf{V_n},$ where the vector of voltages $\mathbf{V_e}$ across the edges can be expressed as  $\mathbf{V_e}=B^T \mathbf{V_n},$ exploiting the fact that the incidence matrix of a directed graph operates as a difference matrix.  

\section{Results}\label{sec:results}
\textit{Note:  In order to maintain readability throughout this section and to avoid overloading the discussions with too many figures, we present our data for $D/L=10.$ The figures corresponding to $D/L=25$ and $50$ can be found in the Appendix~\ref{app:DL2550Plots}.}
 
We start our analysis by comparing the shapes of the percolating clusters in equilibrated isotropic and nematic suspensions. In analogy to how the shape of polymer chains or random walks are quantified, we compute the gyration tensor. In particular, we are interested in the radius of gyration $R_g,$ the asphericity $AS,$ and the aspect ratio $AR$ as defined below: 

\begin{align}
R_g^2 =& \lambda_z^2 + \lambda_y^2 + \lambda_x^2 \\ 
AS =& \lambda_z^2 - \frac{\lambda_y^2+\lambda_x^2}{2}\\
AR =& \sqrt{\frac{\lambda_z^2+0.5(\lambda_y^2+\lambda_x^2)}{\lambda_y^2+\lambda_x^2}}
\end{align}
where $\lambda^2_i$ are the eigenvalues of the gyration tensor.
We first study how the shape of the clusters changes as a function of cluster size on approach to percolation. 

\begin{figure}
\centering
\includegraphics[width=\linewidth]{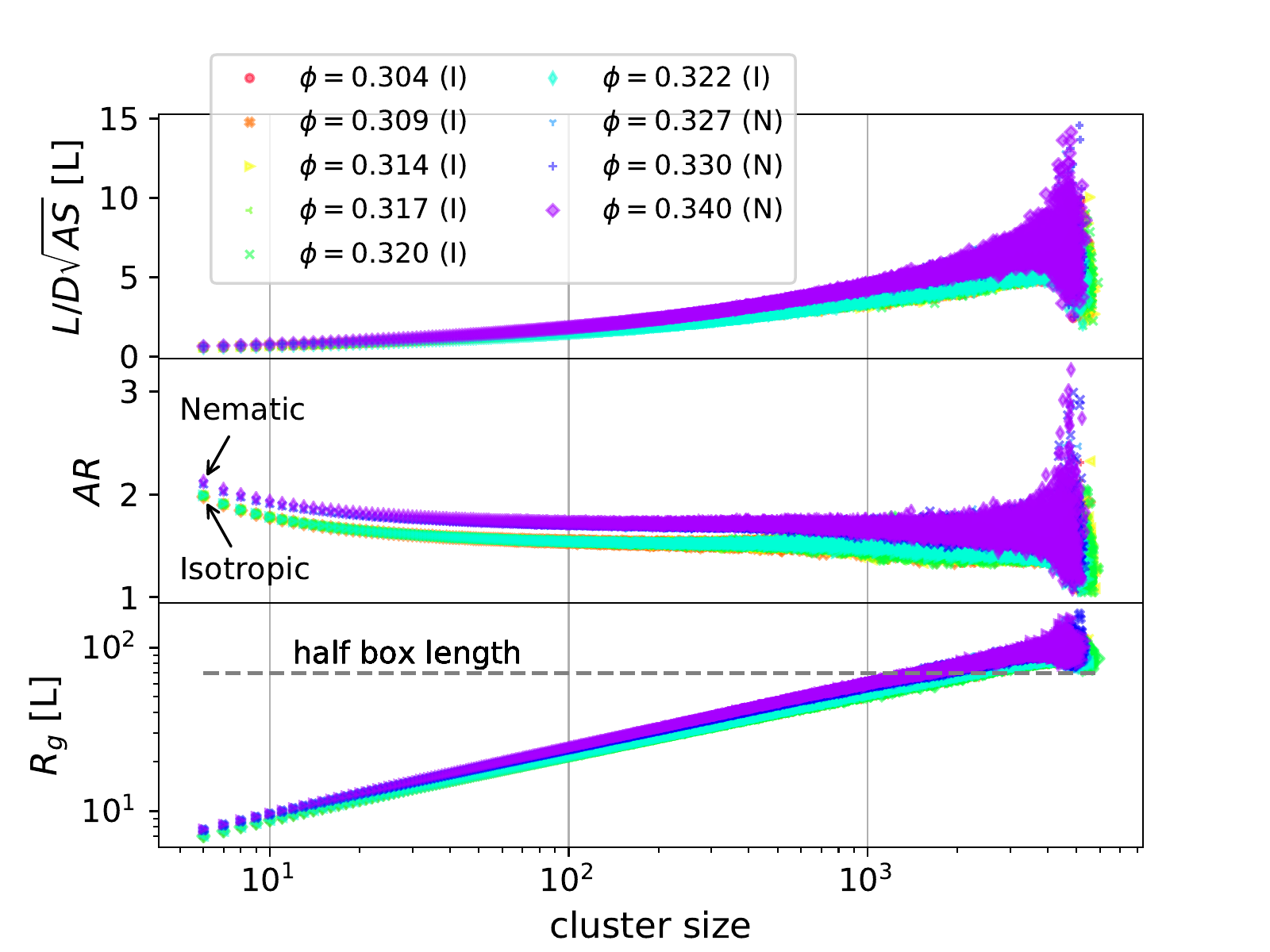}
\caption{Rescaled asphericity (top), aspect ratio (AR, middle) and radius of gyration ($R_g$, bottom) as a function of cluster size for $D/L=10.$ A different marker style and color is used for each volume fraction $\phi$, as shown in the legend, with (I) and (N) denoting isotropic and nematic respectively. In the middle plot, the arrows indicate the branch of data points corresponding to an isotropic and nematic suspension of platelets.}
\label{fig:ASARRgvsnDL10}
\end{figure}

For $D/L=10,$ the combined behaviour of the asphericity (AS) and aspect ratio (AR) shows that while the platelet clusters become less spherically symmetric with their increasing size, their aspect ratio remains constant and only dependent on the degree of alignment of the underlying suspension. In particular, in the top plot of Fig.~\ref{fig:ASARRgvsnDL10}, the asphericity is rescaled to $\sqrt{AS}L/D,$ which offers a more intuitive interpretation, namely, the length difference between the long and the short axes of a cluster expressed in units of shell thickness $L,$ which can thus roughly be interpreted as the difference in particle number along the said axes. 

Moreover, from the behaviour of the cluster aspect ratio as a function of its size (middle, Fig.~\ref{fig:ASARRgvsnDL10}) we notice two different plateau values. Namely, a lower branch of overlapping curves marked ``Isotropic'' corresponding to all simulation volume fractions in the stable \textit{isotropic}, and an upper branch of curves marked ``Nematic'' corresponding to clusters in the stable \textit{nematic} suspensions where we see a small but consistent rise of their respective plateau values with increasing $\phi.$ The latter will become more apparent when we discuss the aspect ratio of the largest cluster as a function of volume fraction.

From the behaviour of the radius of gyration as a function of cluster size as shown in the bottom plot of Fig.~\ref{fig:ASARRgvsnDL10}, we observe a scale independent relation of type $R_g = \alpha n^{1/d_f},$ with the fractal dimension $d_f\approx 2.5$ at percolation, consistent with what is expected for percolating structures in $3D.$ 
We note that all of our above results are consistently obtained for the larger particle aspect ratios $D/L=25, 50,$ and their respective plots can be found in the Appendix~\ref{app:DL2550Plots} (Figs.~\ref{fig:ASARRgvsnDL25} and \ref{fig:ASARRgvsnDL50}). 

\begin{figure}
\centering
\includegraphics[width=\linewidth]{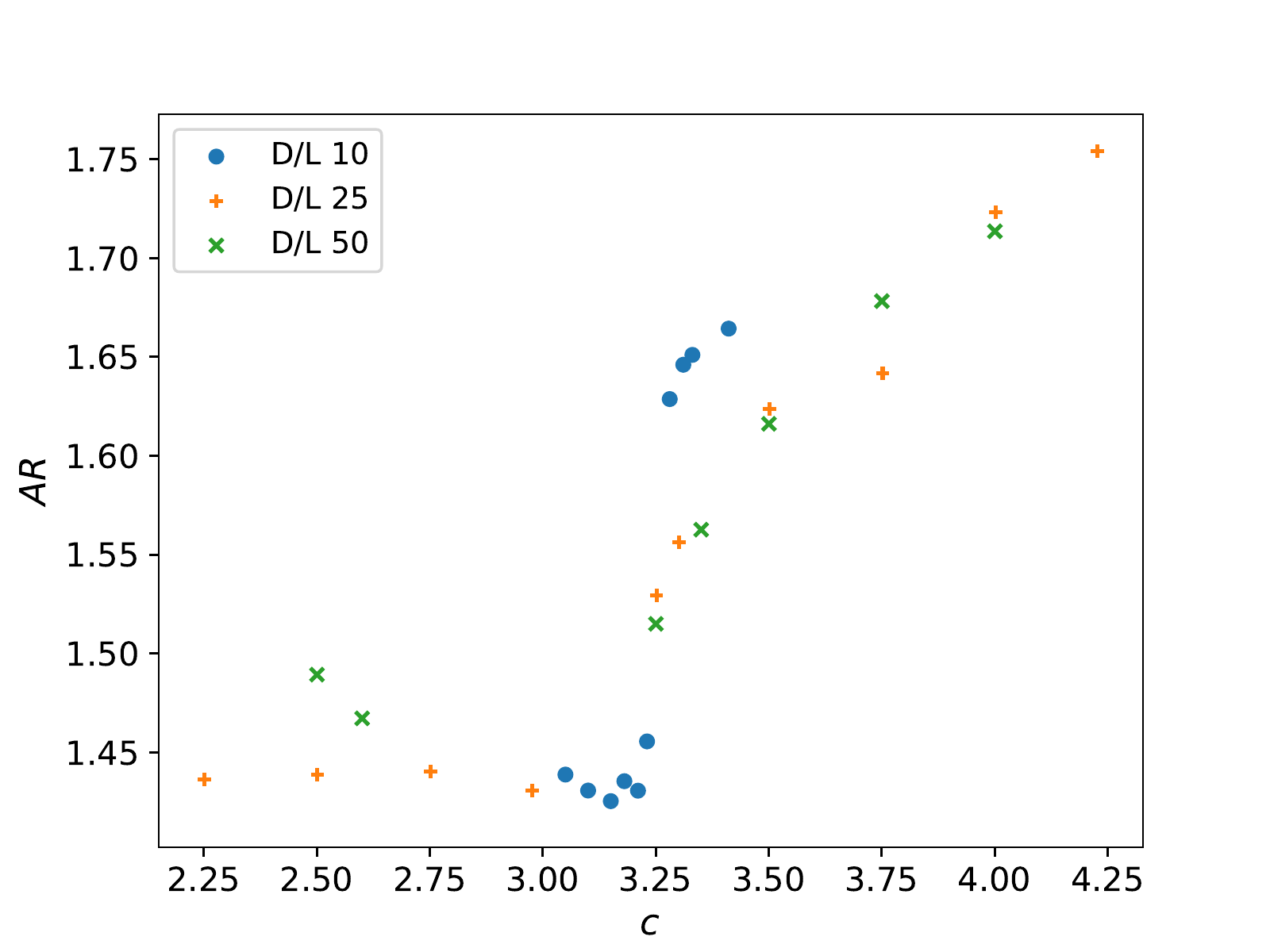}
\caption{Aspect ratio $AR$ of the largest clusters as a function of the rescaled volume fraction $c$ for $D/L=10, 25,$ and $50.$}
\label{fig:ARvseta102550}
\end{figure}

In Fig.~\ref{fig:ARvseta102550}, we have computed the aspect ratio of only the largest clusters, and for convenience, they are plotted as a function of the rescaled volume fraction $c=\phi(L/D-L^3/3D^3)^{-1}$ of the suspension. For all simulated values of $D/L,$ we consistently observe anisotropic cluster shapes with a constant aspect ratio of $AR \approx 1.45$ in the stable isotropic region, which remains weakly dependent on the particle aspect ratio. While in the stable nematic, $AR$ increases monotonically after the jump at the onset of the ordered phase. The latter behaviour already hints at a percolation threshold that decreases with increasing degree of alignment in the suspension. In other words, the more elongated cluster shapes are favourable for percolation as they can more easily become system-spanning.

\begin{figure}
\centering
\includegraphics[width=\linewidth]{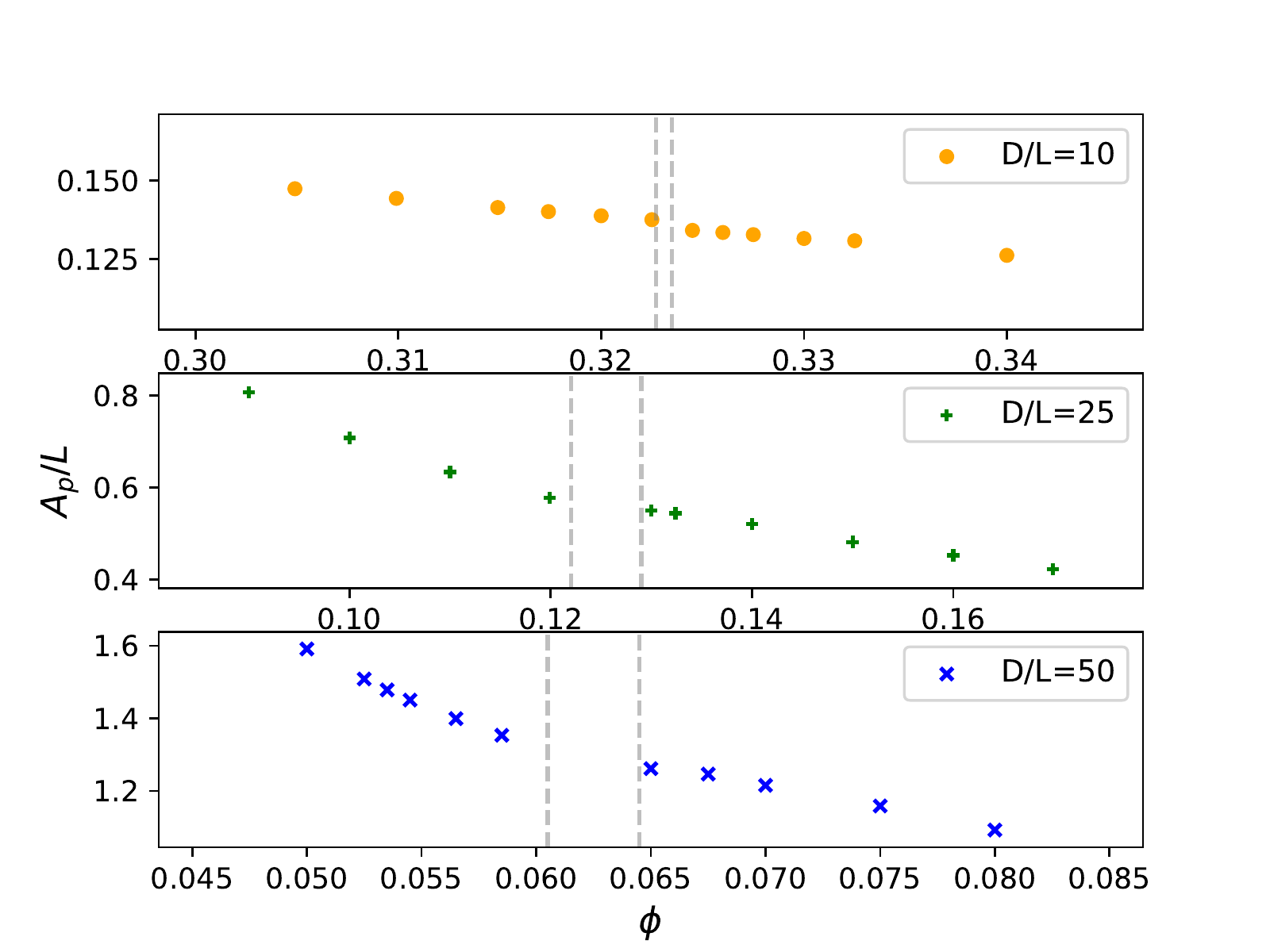}
\caption{Critical shell sizes $A_p/L$ as a function of volume fraction $\phi.$ The plot from top to bottom correspond to $D/L=10, 25$ and $50$ respectively. In each case, the simulation estimates for the I-N coexistence region is indicated by the area between the vertical dashed lines.}
\label{fig:CritShells}
\end{figure}

We now turn to the connectedness percolation thresholds $A_p/L$. As shown in Fig.~\ref{fig:CritShells}, the critical shell size is a monotonically decreasing function of the volume fraction, both in the stable isotropic as well as the nematic phase.  This observation is in agreement with previous MC simulation results\cite{MathewPercolation}. Percolation is always aided by the addition of particles, and particularly in the nematic phase. Unlike the case of rods~\cite{finner2019unusual}, the platelets do not exhibit a re-entrance type of behaviour. 

\begin{figure}
\centering
\includegraphics[width=1.0\linewidth]{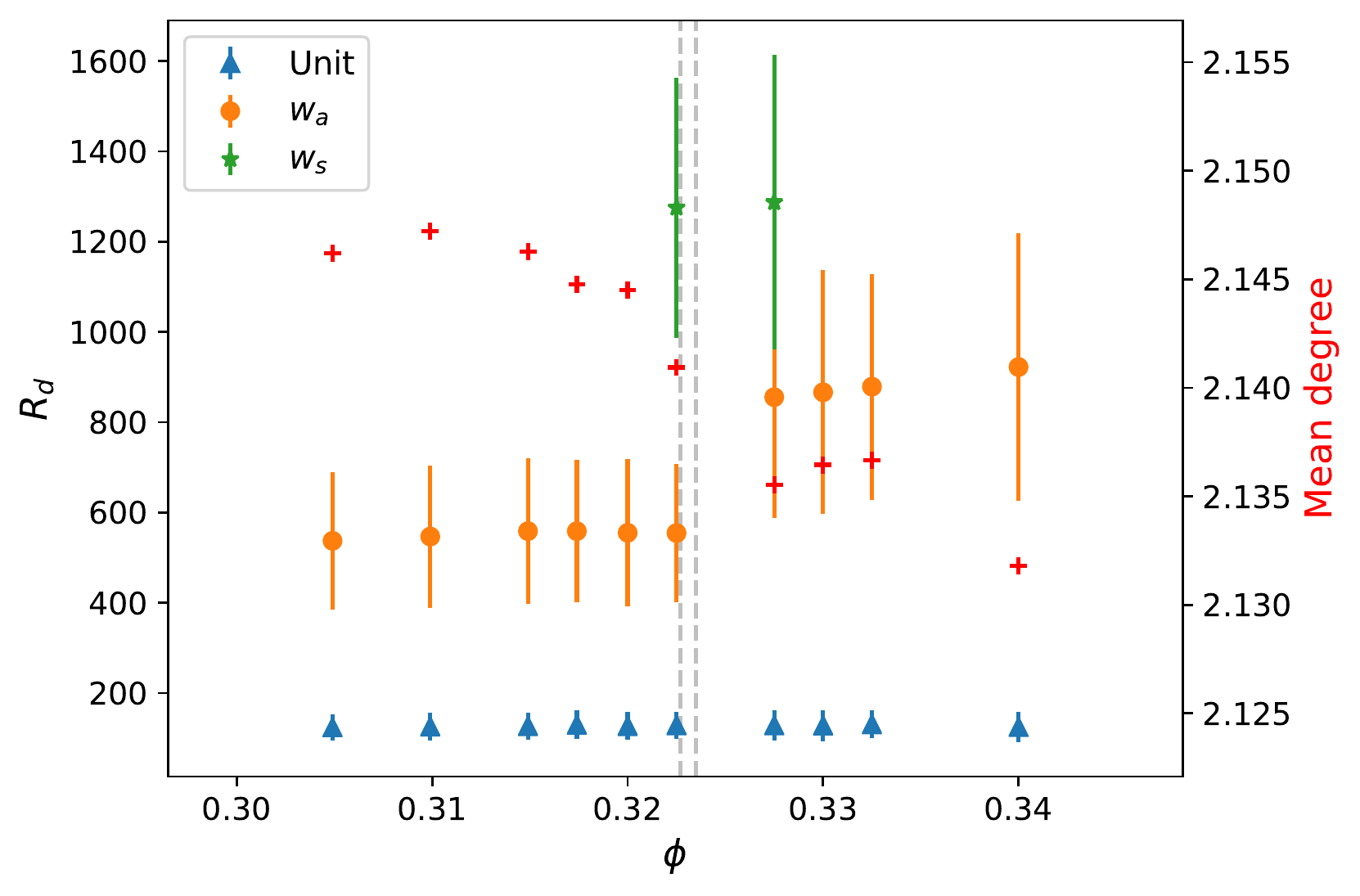}
\caption{Comparing the dependence of the effective resistance between furthest nodes of the network $R_d$ on $\phi$ ($D/L=10,$) according to: unit conductances (labelled \textit{Unit}, triangle, blue), the anisotropic model $w_a$ Eq.~\ref{eq:geomstack} (circle, orange) and the surface-surface overlap model $w_s$ (star, green). The mean node degree for the sampled largest clusters of each $\phi$ is shown on the second Y-axis (plus, red). The dashed vertical lines indicate the simulation estimates of the I-N coexistence region.}
\label{fig:Rdiam10}
\end{figure}

Moreover, the decreasing trend of the percolation threshold appears to be unperturbed by the spontaneous gain of orientational alignment at the onset of the nematic. Again, this is in sharp contrast to the case of rod-like particles, where the number of contacts per rod decreases in the nematic phase due to the increased orientational alignment, which in turn leads to a higher percolation threshold~\cite{finner2019connectivity}. Whereas for platelets, the average number of contacts per particle exhibits a constant trend ($\approx 2.1$), and becomes negligibly ($<1\%$) smaller deep in the nematic, as shown on the 2nd Y-axis of Fig.~\ref{fig:Rdiam10} (Appendix \ref{app:DL2550Plots}: \ref{fig:Rdiam25} and \ref{fig:Rdiam50}) of the following discussions on conductivity.

Having characterized the overall shape properties of the largest clusters, we study next their conductivity and current flow properties at percolation. Throughout the following part, our results are shown in terms of the inverse conductivity, i.e., the electrical resistance.  For the comparison of the resistance behavior across the intrinsic isotropic and nematic phases, we start by performing simple two-point resistance measurements across the diameter of the platelet network, i.e., the electrodes are placed at the two nodes furthest apart according to the unweighted graph distance. Then in order to ensure that our observations are not merely an artifact of the choice of measurement points, we also compute the normalized Kirchhoff indices Eq.~\ref{eq:kirchindex} (often referred to as \textit{network criticality}) of the networks as a function of volume fraction:
\begin{equation}\label{eq:kirchindex}
\hat{\tau} = \frac{2}{N(N-1)} \sum_{i=1}^{N} \sum_{j=i+1}^{N} R_{ij}
\end{equation}
The chosen networks, which comprise our averaging set, correspond to $500-1000$ independent realisations of the largest cluster of equilibrated and percolating suspensions of hard platelets.

\begin{figure}
\centering
\includegraphics[width=1.0\linewidth]{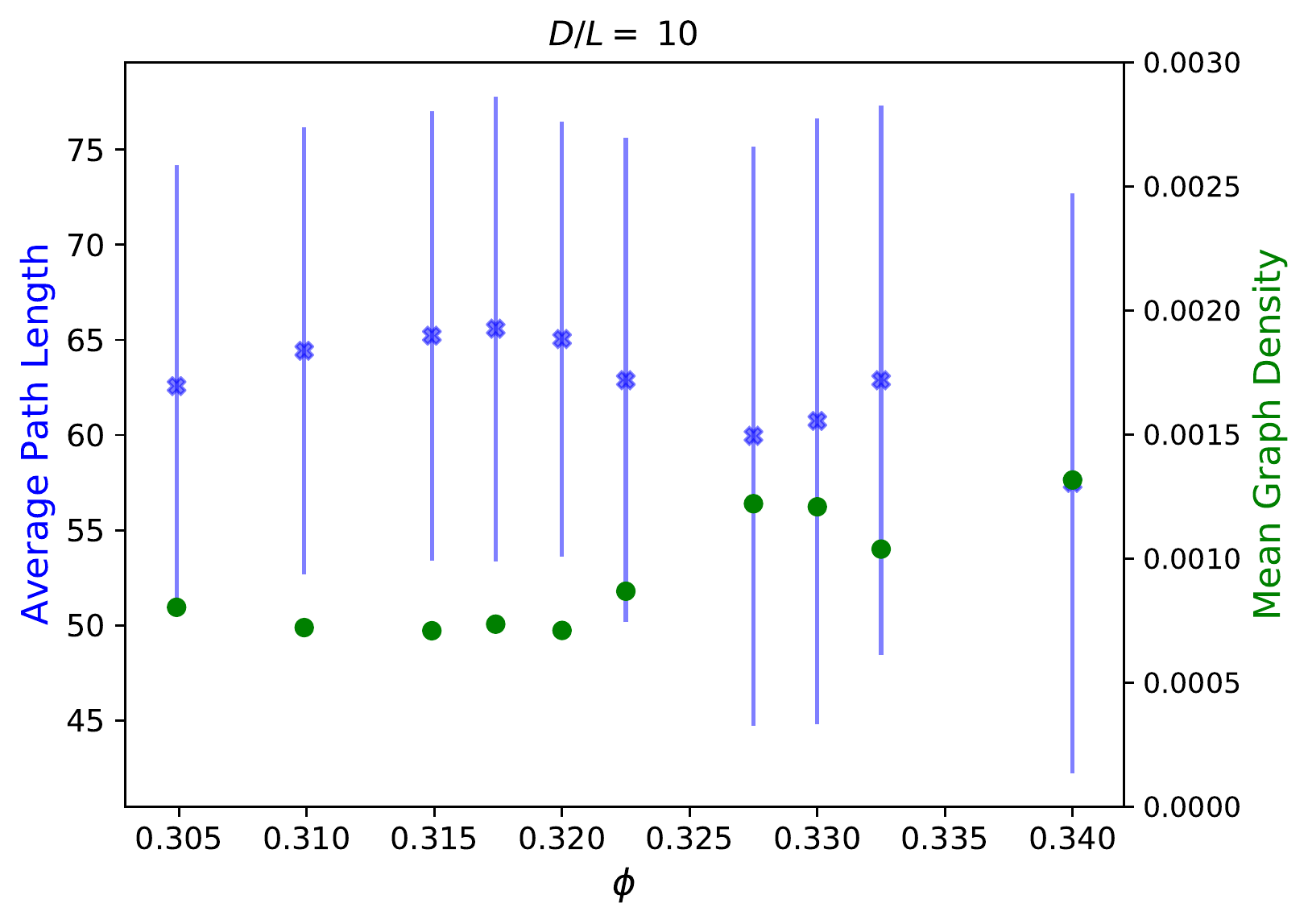}
\caption{Mean average path length (x, blue) and mean graph density (circle, green) of tunneling networks as a function of volume fraction $\phi,$ for $D/L=10.$
\label{fig:aveplDensity10}}
\end{figure}

\begin{figure}
\centering
\includegraphics[width=1.0\linewidth]{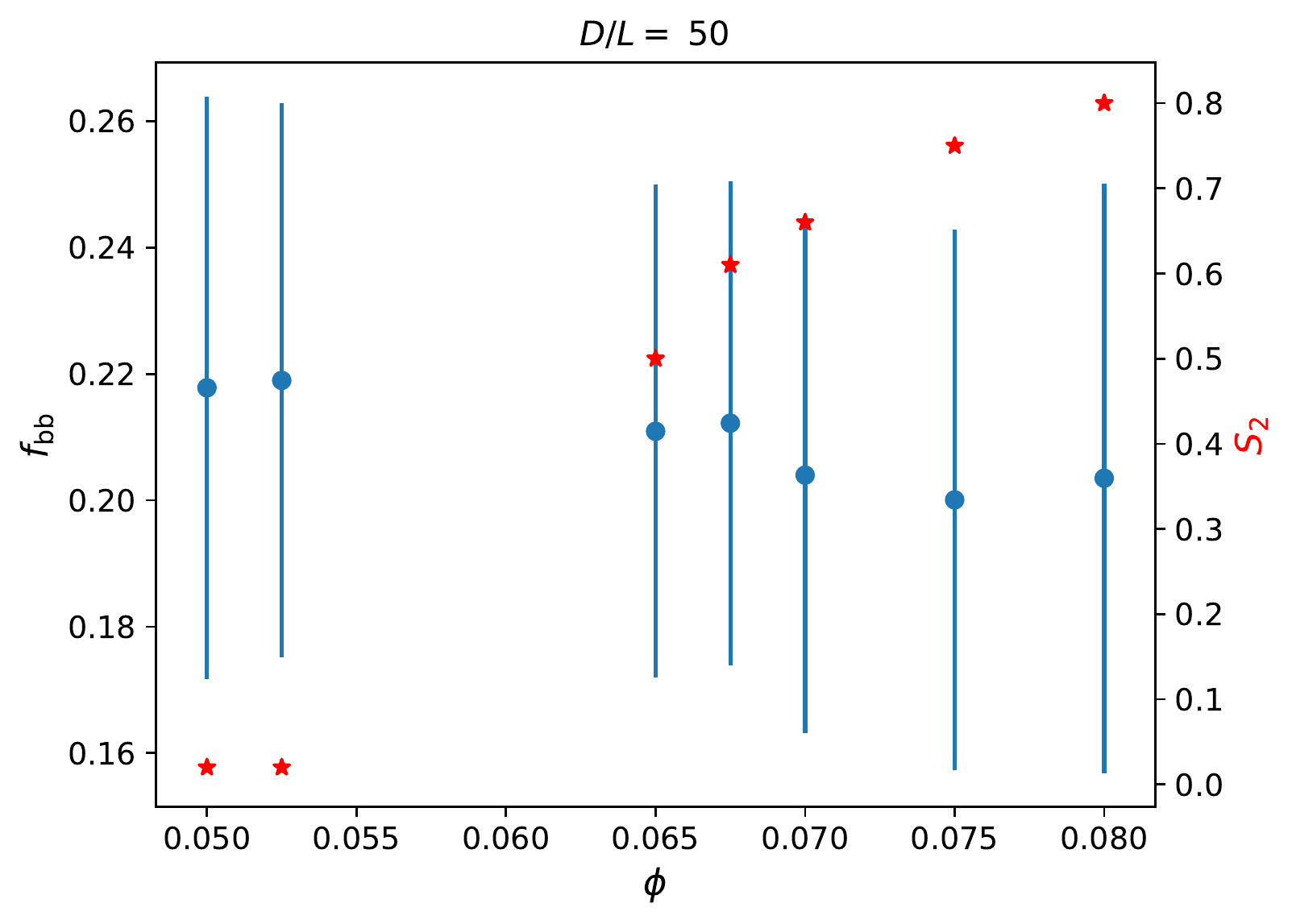}
\caption{Fraction $f_{\rm{bb}}$ (circle, blue) of the network nodes comprising the conductive backbone as a function of volume fraction $\phi$ for $D/L=50.$ The backbone here is defined as the set of current carrying bonds for a unit current inserted and extracted at the diameter nodes of the network. On the second Y-axis, the nematic order parameter $S_2$ (star, red) of the network's underlying suspension is plotted.}
\label{fig:bbFracDiam50}
\end{figure}

\begin{figure}
\centering
\includegraphics[width=1.0\linewidth]{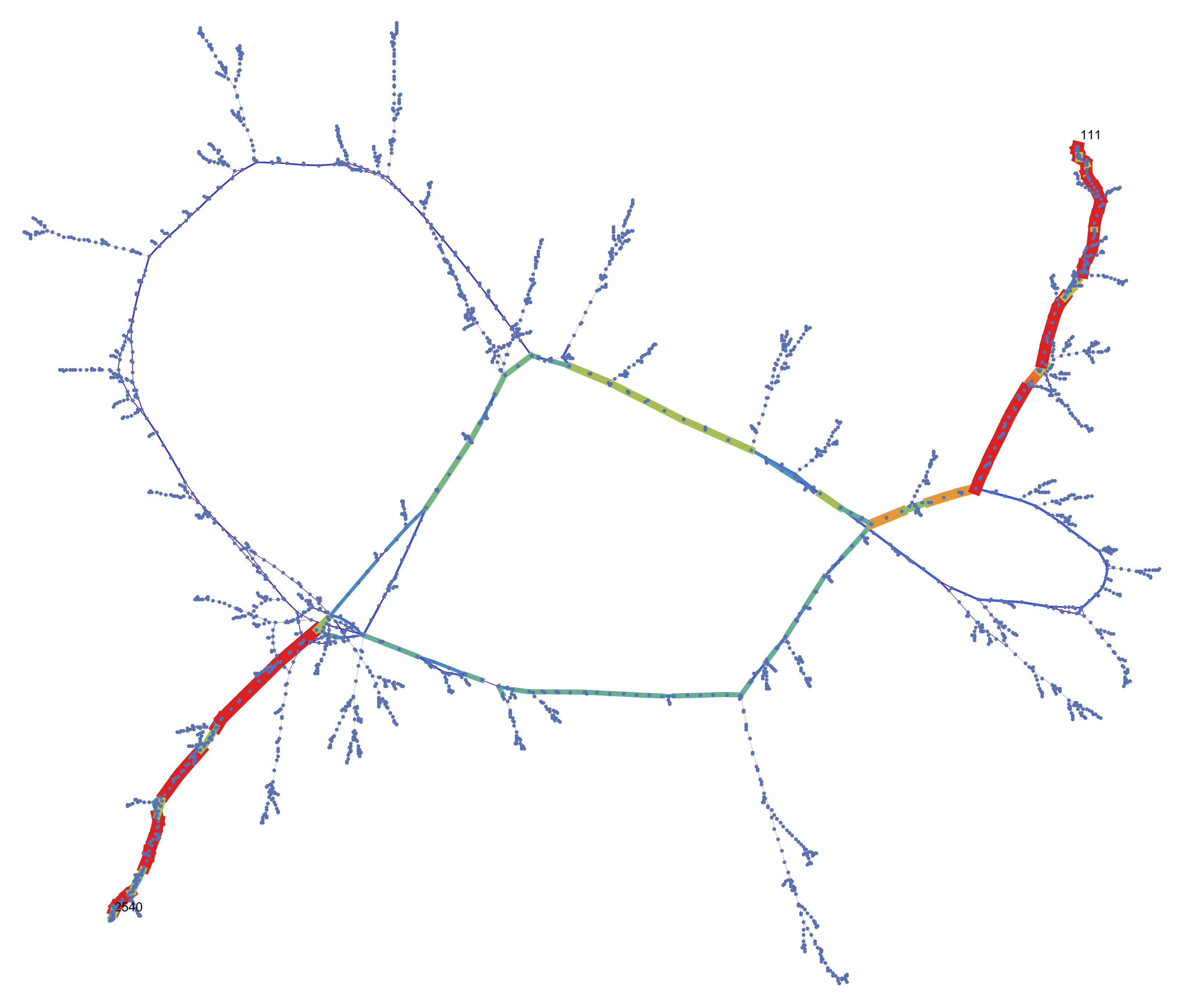}
\caption{A network and its backbone (source node: $2540$, sink node: $111$) highlighted for $D/L=50$ and $\phi=0.0525.$ Bond thickness and color denote the amplitude of the current running through it.}
\label{fig:backbonevisual}
\end{figure}

The results for $D/L=10$ are shown in Fig.~\ref{fig:Rdiam10} (Appendix~\ref{app:DL2550Plots}, Figs.~\ref{fig:Rdiam25} and \ref{fig:Rdiam50} for $D/L=25, 50$). On the one hand, in the case of simple networks (unit conductances), we observe a constant resistance behavior: the simple act of adding more platelets to the suspension does not lead to an improved conductivity of the corresponding network. Moreover, as shown in Fig.~\ref{fig:bbFracDiam50} for $D/L=50,$ the fraction of the networks comprising the conductive backbone is nearly unaffected by the addition of particles and is even lowered in the nematic, i.e., the percolating clusters become more populated by dangling ends. The latter points clearly highlight that the structural changes of the tunneling networks induced by the spontaneous alignment of the platelets do not improve their current flow properties. For illustration, a cluster with its highlighted backbone is shown in Fig.~\ref{fig:backbonevisual}.
These observations are consistent with other topological aspects of the networks, as shown in Fig.~\ref{fig:aveplDensity10} (Appendix~\ref{app:DL2550Plots}, Figs.~\ref{fig:aveplDensity25} and \ref{fig:aveplDensity50}), where both the graph density and average path length remain nearly constant as a function of volume fraction, and are unaffected by the IN transition.
Furthermore, the observed high average path length values ($\approx 45-60$) and very low graph densities ($ < 0.003$) suggest that the networks are very sparsely connected.

On the other hand, according to the anisotropic model of Eq.~\ref{eq:geomstack} which weighs conductances by explicitly accounting for the relative orientation of connected platelets, we consistently observe for all aspect ratios ($10,25,$ and $50$) a monotonically increasing resistance across the IN transition. And the same trend persists throughout the respective stable phases. 

The increased resistance can be explained by considering that in spite of the increased alignment of platelets in the nematic, their center of mass positions remain weakly correlated. Moreover, the elongated shape of the percolating clusters suggests that the contiguously connected pairs of platelets that comprise the cluster are more likely to have their centers mis-aligned. These observations imply that having pairs of platelets with overlapping shells that are in an aligned and centered configuration (Fig.~\ref{fig:snape}) do not become more probable in the nematic phase. Instead, they are more likely to be side-by-side connected (Fig.~\ref{fig:snapa}, having the lowest conductance) or have partial overlap (Fig.~\ref{fig:snapb}). In contrast, in the isotropic phase, inherently due to the disordered orientations, connections of the type (Fig.~\ref{fig:snapc}) are more likely to occur, which correspond to higher conductance values compared to the two former cases. 

\begin{figure}
\centering
\includegraphics[width=1.0\linewidth]{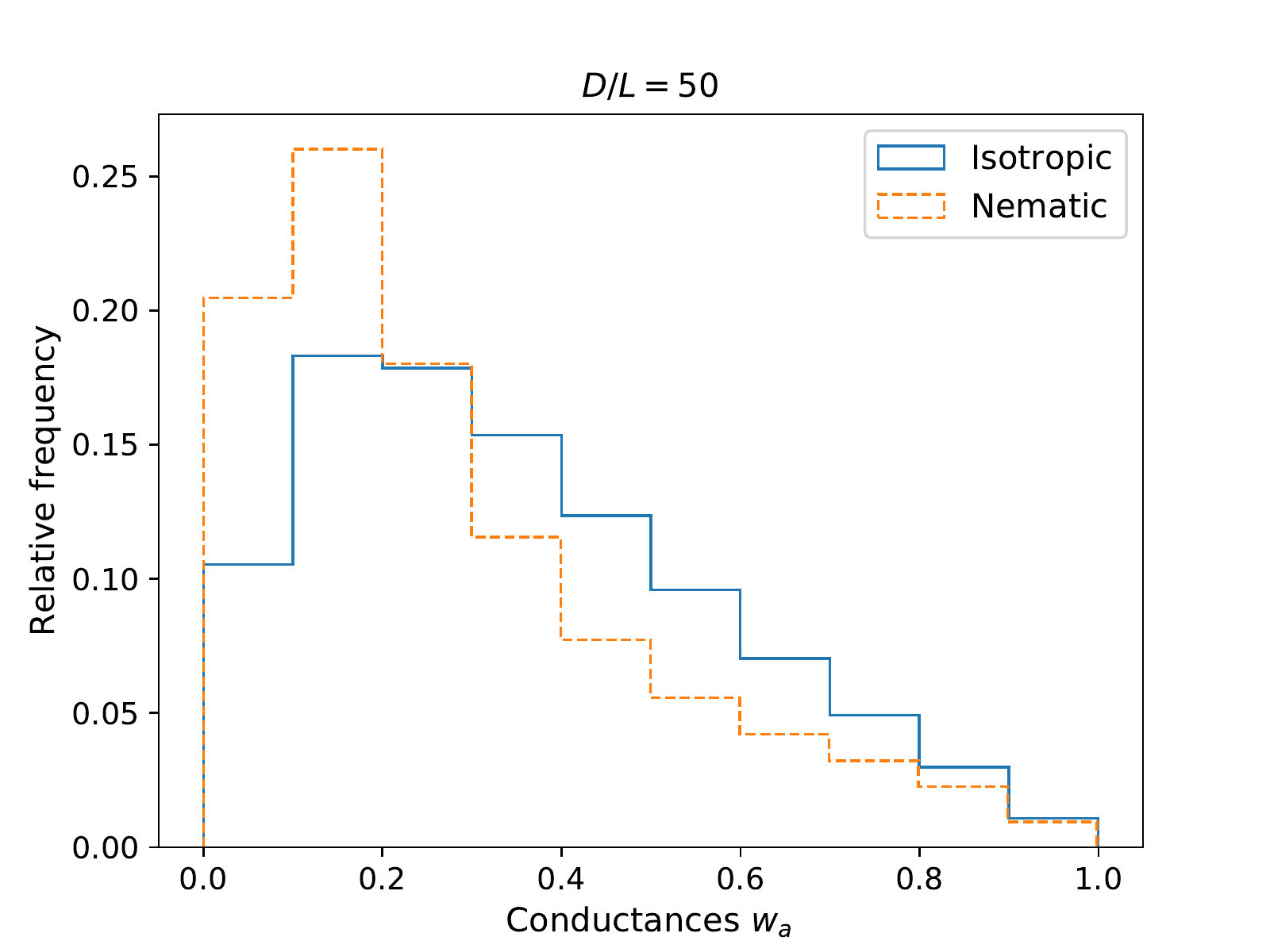}
\caption{Distribution of conductances $w_a$ for tunneling networks of $D/L=50.$ The  dashed orange histogram corresponds to networks selected from $200$ nematic suspensions of $\phi=0.07.$ Similarly, the solid blue histogram corresponds to those selected from $200$ isotropic suspensions of $\phi=0.0525.$}
\label{fig:histDL50wa}
\end{figure}

This argument is corroborated by the plot shown in Fig.~\ref{fig:histDL50wa} for $D/L=50$: the orange (dashed) and blue (solid) histograms show the conductances computed according to Eq.~\ref{eq:geomstack}, for networks drawn from independent realisations of nematic ($\phi=0.07$) and isotropic suspensions ($\phi=0.0525$). The nematic histogram displays a noticeable shift to smaller conductances, with a more pronounced peak at $w_a\approx 0.2$, which indeed corresponds to configurations of type shown in Fig.~\ref{fig:snapb}. 

Finally, for the second anisotropic model $w_s,$ which assigns conductances proportionally to the mean surface-surface overlap of the platelets, we again do not observe an enhancement of the conductivity at the IN transition. (Unfortunately, due to CPU time limitations and the highly expensive surface-surface overlap computations, our comparison with the other models cannot be performed on an equal statistical footing. We have only considered two points in volume fraction and averaged the conductivity only over $100$ realisations of the networks.)

Before discussing our findings for the Kirchhoff indices of the tunneling networks, it is important to briefly elucidate on how they provide a useful geometric measure for comparing the connectivity of networks in view of their current flow properties. On the one hand, a high Kirchhoff index or total resistance is a simple indicator for the network being poorly wired for accommodating current flows. 

On the other hand, on a more intuitive level, it also acts as a measure for how structurally robust a network is. This can be more easily seen by first noting that $\tau$ (here not normalized) can be expressed as~\cite{ellens2011effective} $\tau = N\operatorname{Tr}(K^+)=N\sum_{i=2}^N \lambda_i^{-1}$ where $K^+$ and $\lambda_i$'s are the Moore-Penrose pseudo-inverse and eigenvalues of the Laplacian respectively. The inverse eigenvalues of $K$ can be interpreted as the topological centrality~\cite{RANJAN20133833} of the nodes. Thus, a lower Kirchhoff index is indicative of a more compactly connected structure, having a lower average node centrality. The latter naturally relates to the network robustness, since having a lower average node centrality means the current flow properties of the network are not sensitive to a small set of nodes, whose removal would drastically alter the current flow efficiency across the network.

\begin{figure}
\centering
\includegraphics[width=1.0\linewidth]{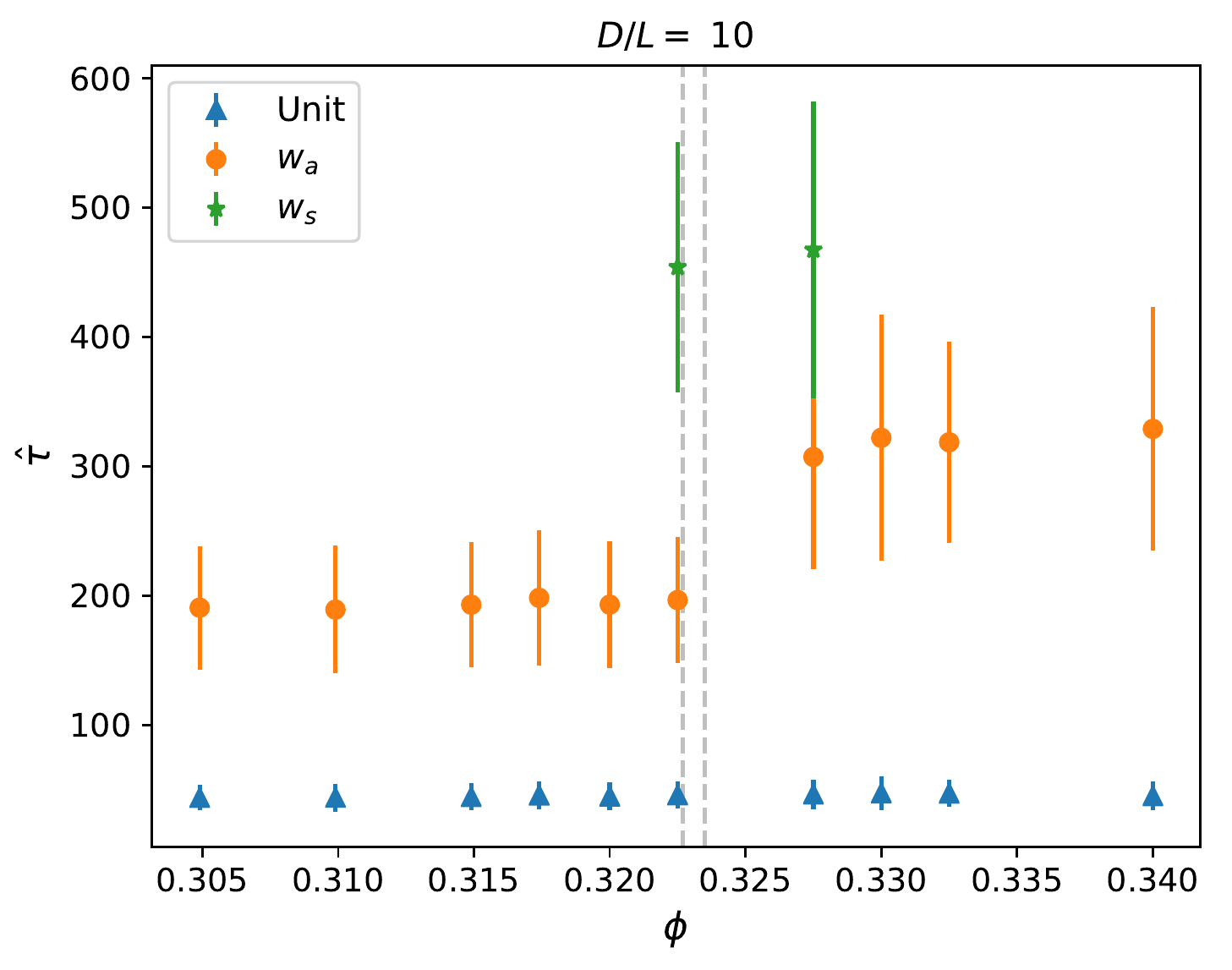}
\caption{Normalized Kirchhoff index of the tunneling networks (for $D/L=10$) as a function of volume fraction for unit conductances (labelled \textit{Unit}, triangle, blue), for the anisotropic model $w_a$ Eq.~\ref{eq:geomstack} (circle, orange), and for the surface-surface overlap model $w_s$ (star, green).}
\label{fig:taudegree10}
\end{figure}

In Fig.~\ref{fig:taudegree10}, (Appendix~\ref{app:DL2550Plots}: Figs.~\ref{fig:taudegree25} and \ref{fig:taudegree50}) we show the results obtained for the normalized Kirchhoff indices of the networks. In agreement with the measurements made at the furthest two nodes of the network, we consistently observe that $\hat{\tau}$ remains constant for the simple network model, i.e., the addition of platelets does not enhance its robustness. Instead, and particularly for the anisotropic conductance model, $\hat{\tau}$ increases monotonically as a function of volume fraction. Thus suggesting, that the network conductivity is weakened both at the onset of the IN transition and deeper in the stable nematic. These observations agree with the finding, that the structural variation of the tunneling networks induced by either the addition of platelets or their spontaneous alignment, does not lead to an enhancement of the conductivity.

\section{Conclusions}\label{sec:summary}
In summary, we observe that in both cases of orientationally ordered and disordered phases the clusters are anisotropic. While the aspect ratio of the clusters remains constant in the isotropic phase, deep in the nematic phase the platelets  form more elongated structures. The increased elongation in the nematic suggests platelet clusters can be used for their anisotropic transport properties in the context of nanocomposites. Similarly favourable for the latter application, is their percolation behaviour: the threshold $A_p/L$ decreases monotonically with volume fraction and thus, percolation can never be lost with addition of platelets. 

However, in sharp contrast to the aforementioned aspects of platelets which support their use in nanocomposites, the conductivity properties of resistor networks formed by the platelets show no signs of enhancement, neither with the addition of more particles, nor with the gained orientational alignment in the nematic. In fact, when explicitly taking into account the local structure in the assignment of conductances, the conductivity worsens in the nematic. This can be explained by the fact that the increased alignment also leads to a more elongated spatial distribution of the platelets, which in turn implies the surface surface overlap between connected pairs does not increase, despite being orientationally correlated. 
The latter observations are also tightly connected to the fact that the topology of the platelet network remains mostly unaffected by changes in volume fraction and liquid-crystalline phase of the suspension. 

The conductance models studied in this work provide an insight into how the conductivity properties of platelet networks are affected under varying volume fraction and intrinsic alignment conditions.
Therefore, an important line of continuation for future work remains to investigate the network conductivity also under an explicit tunneling-based conductance model, that would, similar to the well-studied case of rods, account both for the tunneling decay with shortest surface to surface distance, and its dependence on the relative orientation of the particles.

\begin{acknowledgments}
The authors acknowledge support by the state of Baden-W\"urttemberg through bwHPC
and the German Research Foundation (DFG) through grants no INST 39/963-1 FUGG and SCHI 853/4-1 648689.
\end{acknowledgments}

\appendix

\section{Remaining plots for $D/L=25$ and $50$}\label{app:DL2550Plots}

\begin{figure}
\centering
\includegraphics[width=\linewidth]{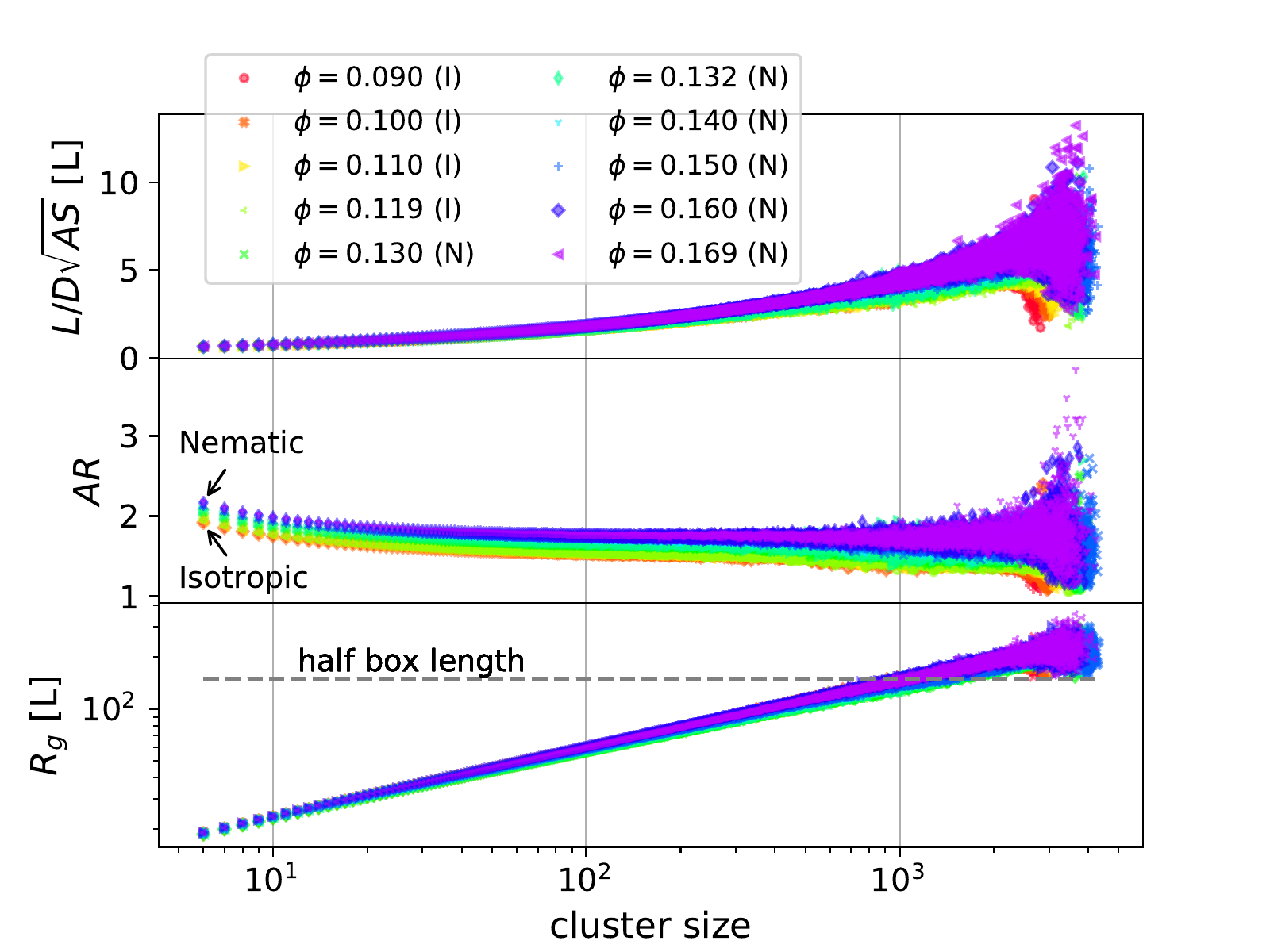}
\caption{Rescaled asphericity (top), aspect ratio (AR, middle) and radius of gyration ($R_g$, bottom) as a function of cluster size for $D/L=25.$ A different marker style and color is used for each volume fraction $\phi$, as shown in the legend, with (I) and (N) denoting isotropic and nematic respectively. In the middle plot, the arrows indicate the branch of data points corresponding to an isotropic and nematic suspension of platelets.}
\label{fig:ASARRgvsnDL25}
\end{figure}

\begin{figure}
\centering
\includegraphics[width=\linewidth]{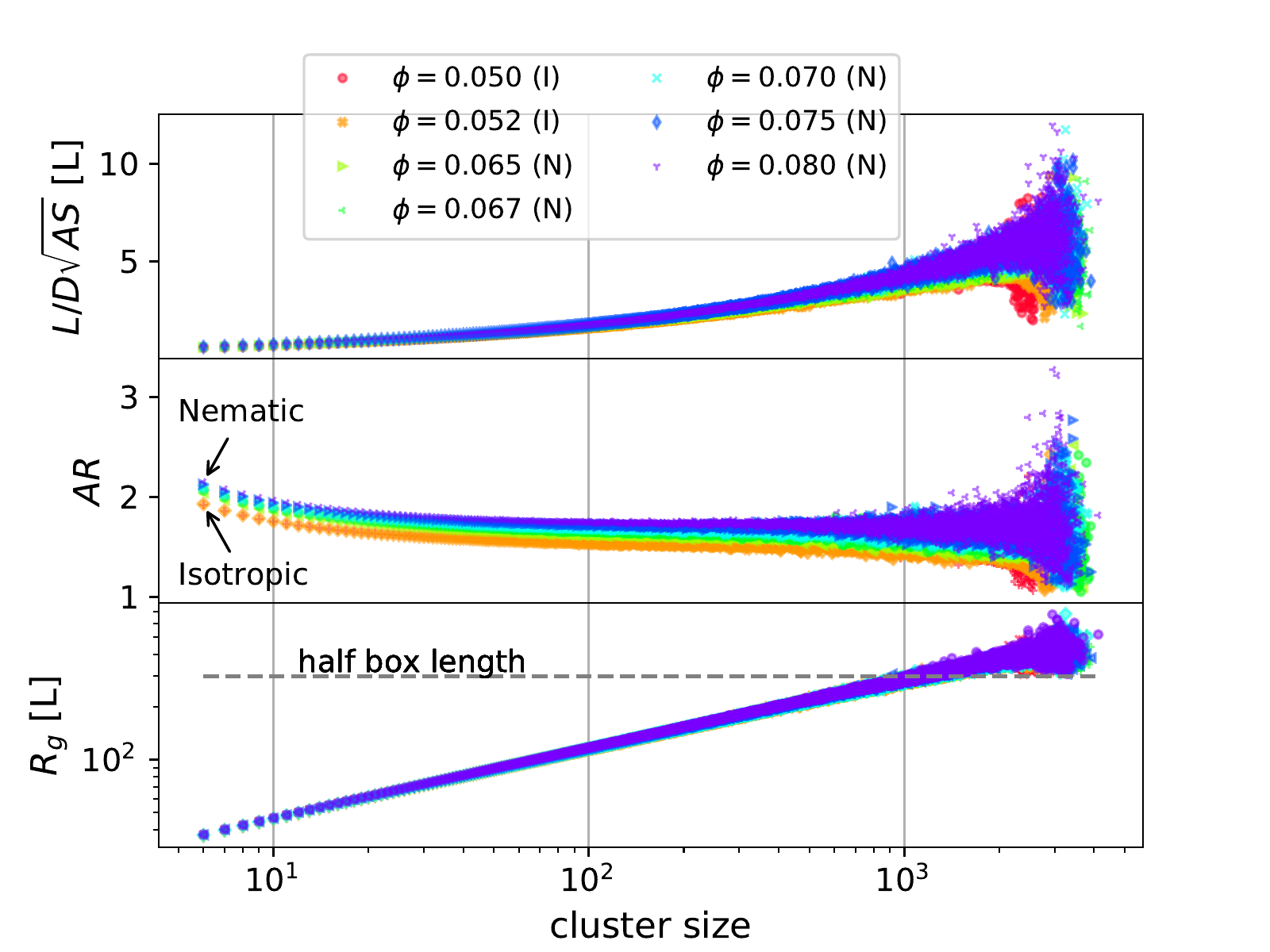}
\caption{Rescaled asphericity (top), aspect ratio (AR, middle) and radius of gyration ($R_g$, bottom) as a function of cluster size for $D/L=50.$ A different marker style and color is used for each volume fraction $\phi$, as shown in the legend, with (I) and (N) denoting isotropic and nematic respectively. In the middle plot, the arrows indicate the branch of data points corresponding to an isotropic and nematic suspension of platelets.}
\label{fig:ASARRgvsnDL50}
\end{figure}

\begin{figure}
\centering
\includegraphics[width=1.0\linewidth]{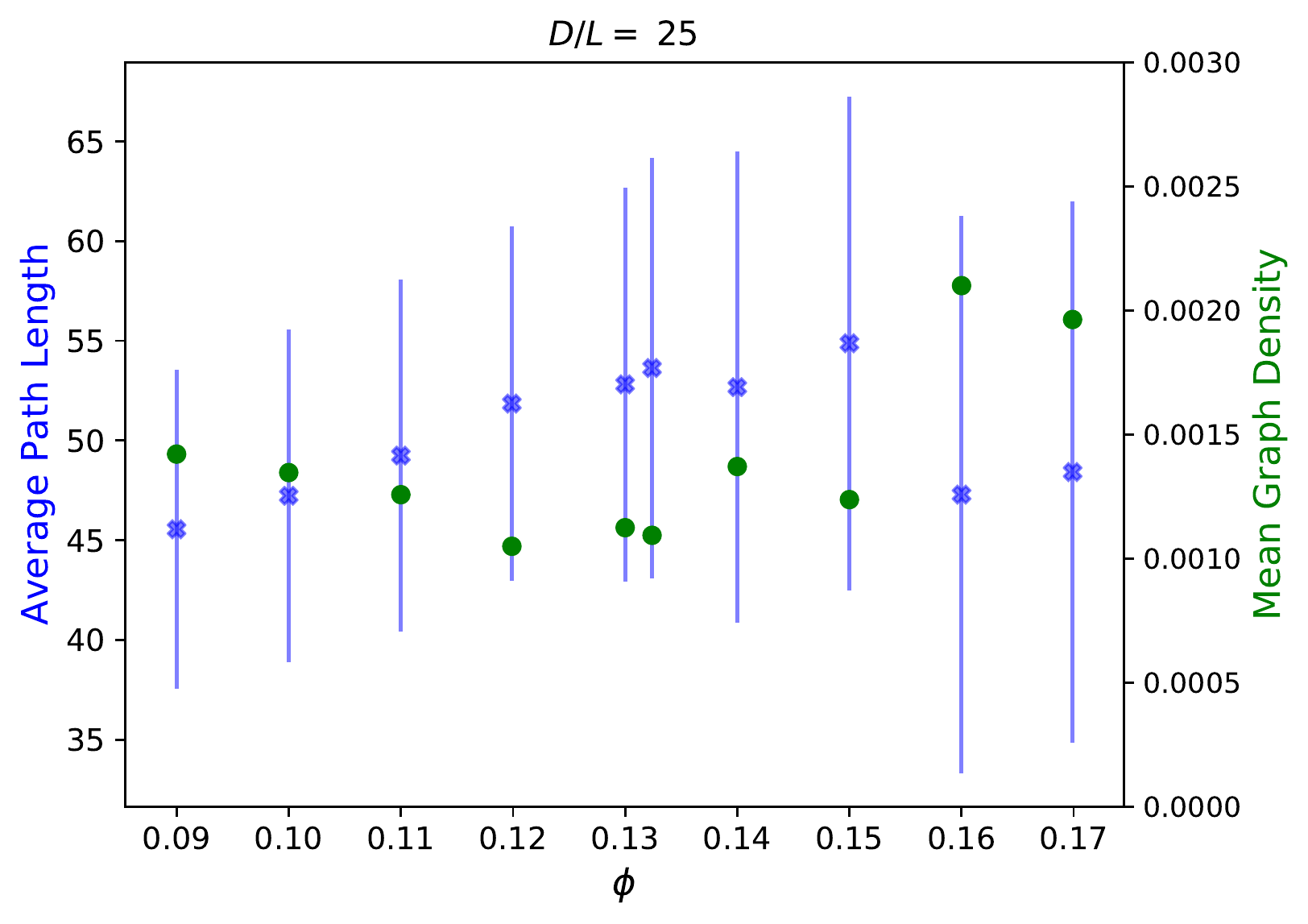}
\caption{Mean average path length (x, blue) and mean graph density (circle, green) of tunneling networks as a function of volume fraction $\phi,$ for $D/L=25.$
\label{fig:aveplDensity25}}
\end{figure}

\begin{figure}
\centering
\includegraphics[width=1.0\linewidth]{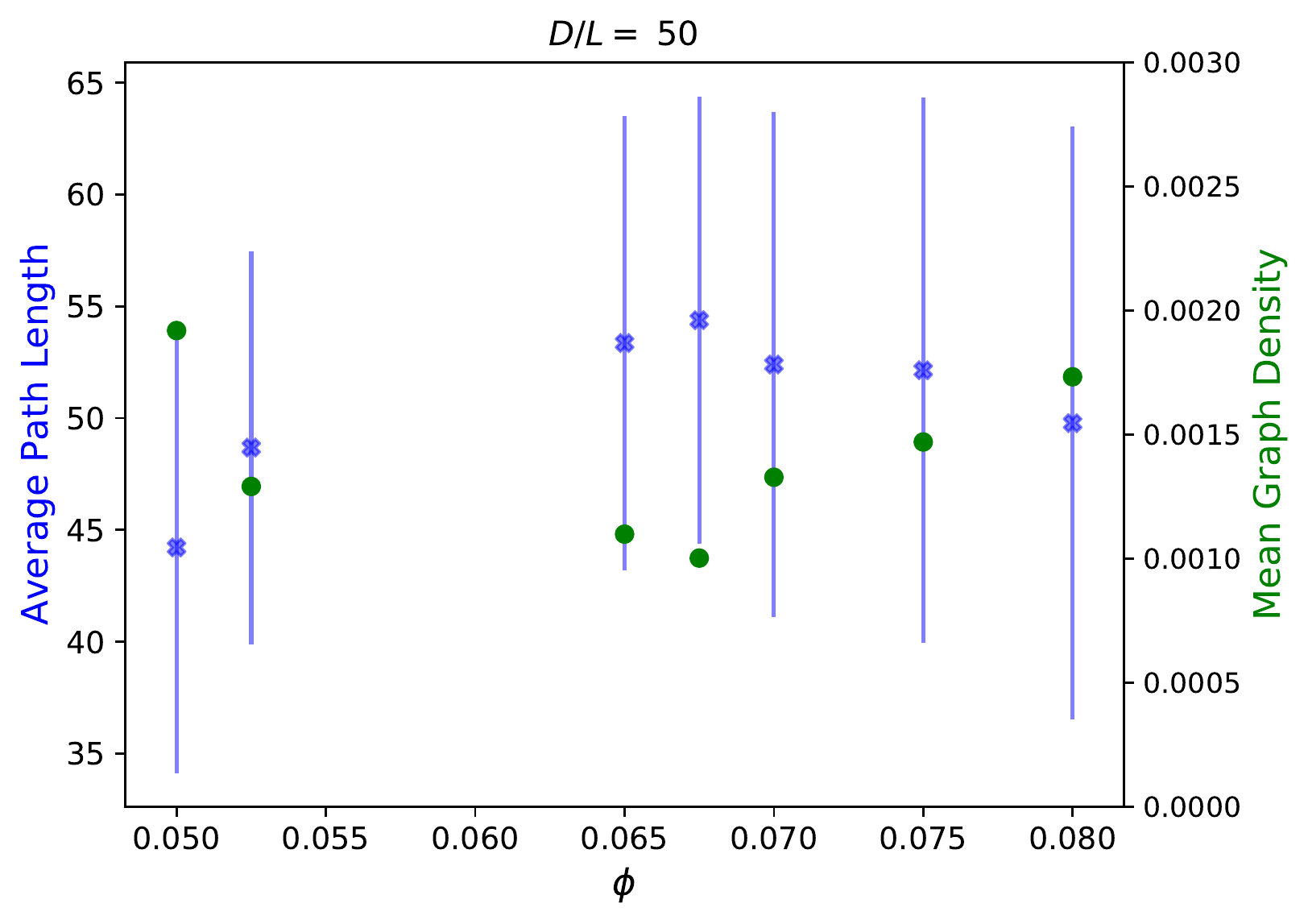}
\caption{Mean average path length (x, blue) and mean graph density (circle, green) of tunneling networks as a function of volume fraction $\phi,$ for $D/L=50.$
\label{fig:aveplDensity50}}
\end{figure}

\begin{figure}
\centering
\includegraphics[width=1.0\linewidth]{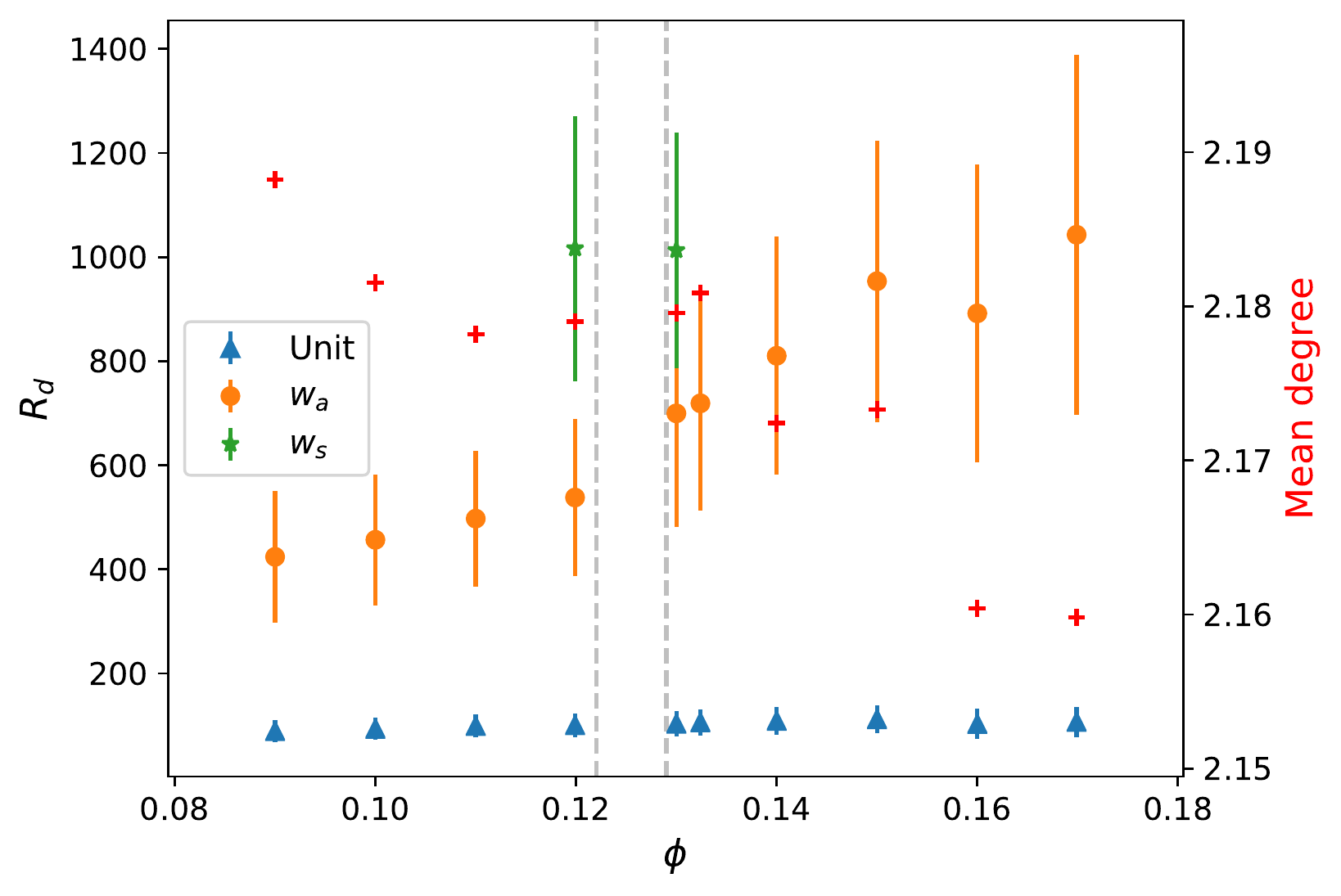}
\caption{Comparing the dependence of the effective resistance between furthest nodes of the network $R_d$ on $\phi,$ and for $D/L=25,$ according to: unit conductances (labelled \textit{Unit}, triangle, blue), the anisotropic model $w_a$ Eq.~\ref{eq:geomstack} (circle, orange) and the surface-surface overlap model $w_s$ (star, green). The mean node degree for the sampled largest clusters of each $\phi$ is shown on the second Y-axis (plus, red). The dashed vertical lines indicate the simulation estimates of the I-N coexistence region.}
\label{fig:Rdiam25}
\end{figure}

\begin{figure}
\centering
\includegraphics[width=1.0\linewidth]{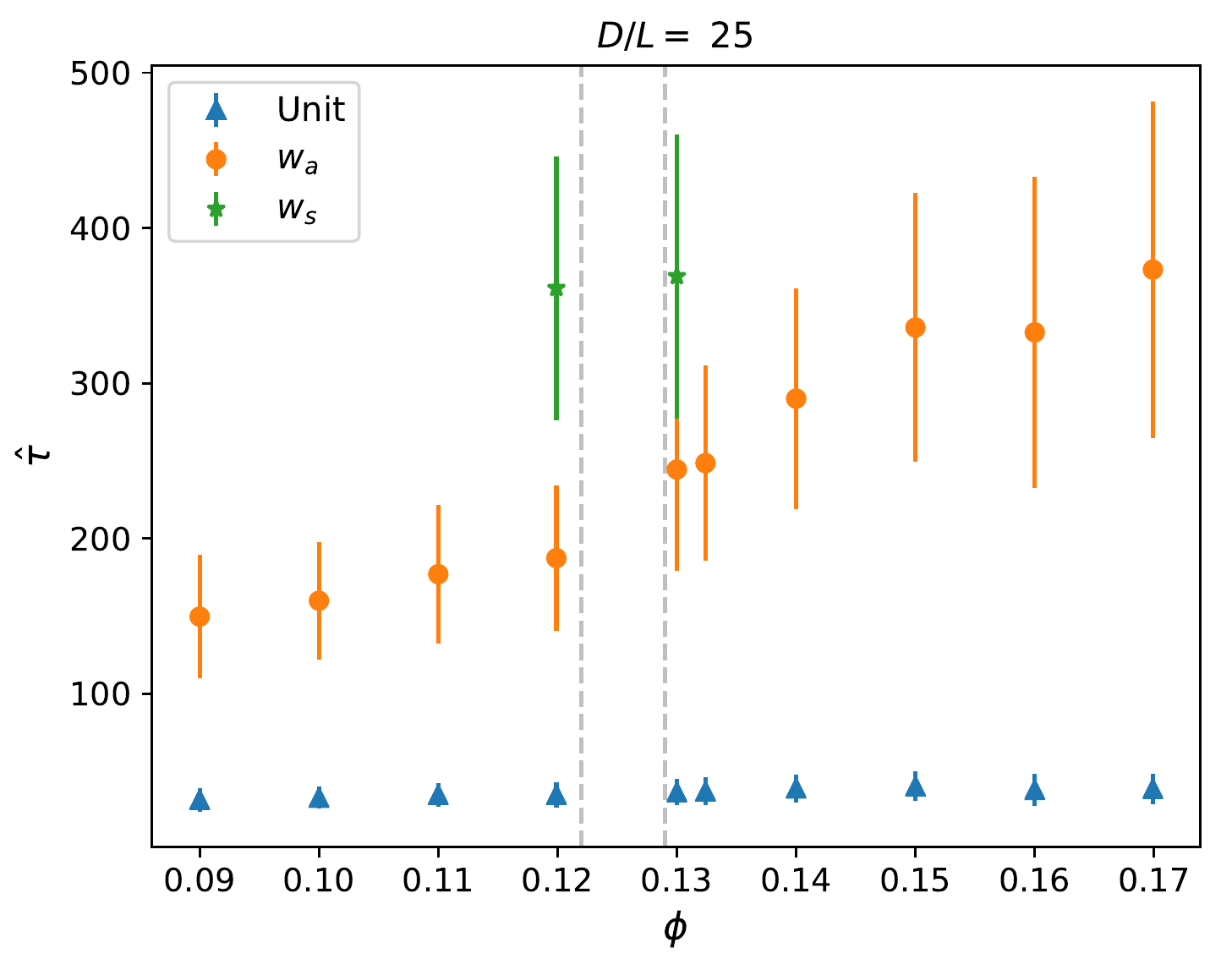}
\caption{Normalized Kirchhoff index of the tunneling networks (for $D/L=25$) as a function of volume fraction for unit conductances (labelled \textit{Unit}, triangle, blue), for the anisotropic model $w_a$ Eq.~\ref{eq:geomstack} (circle, orange), and for the surface-surface overlap model $w_s$ (star, green).}
\label{fig:taudegree25}
\end{figure}

\begin{figure}
\centering
\includegraphics[width=1.0\linewidth]{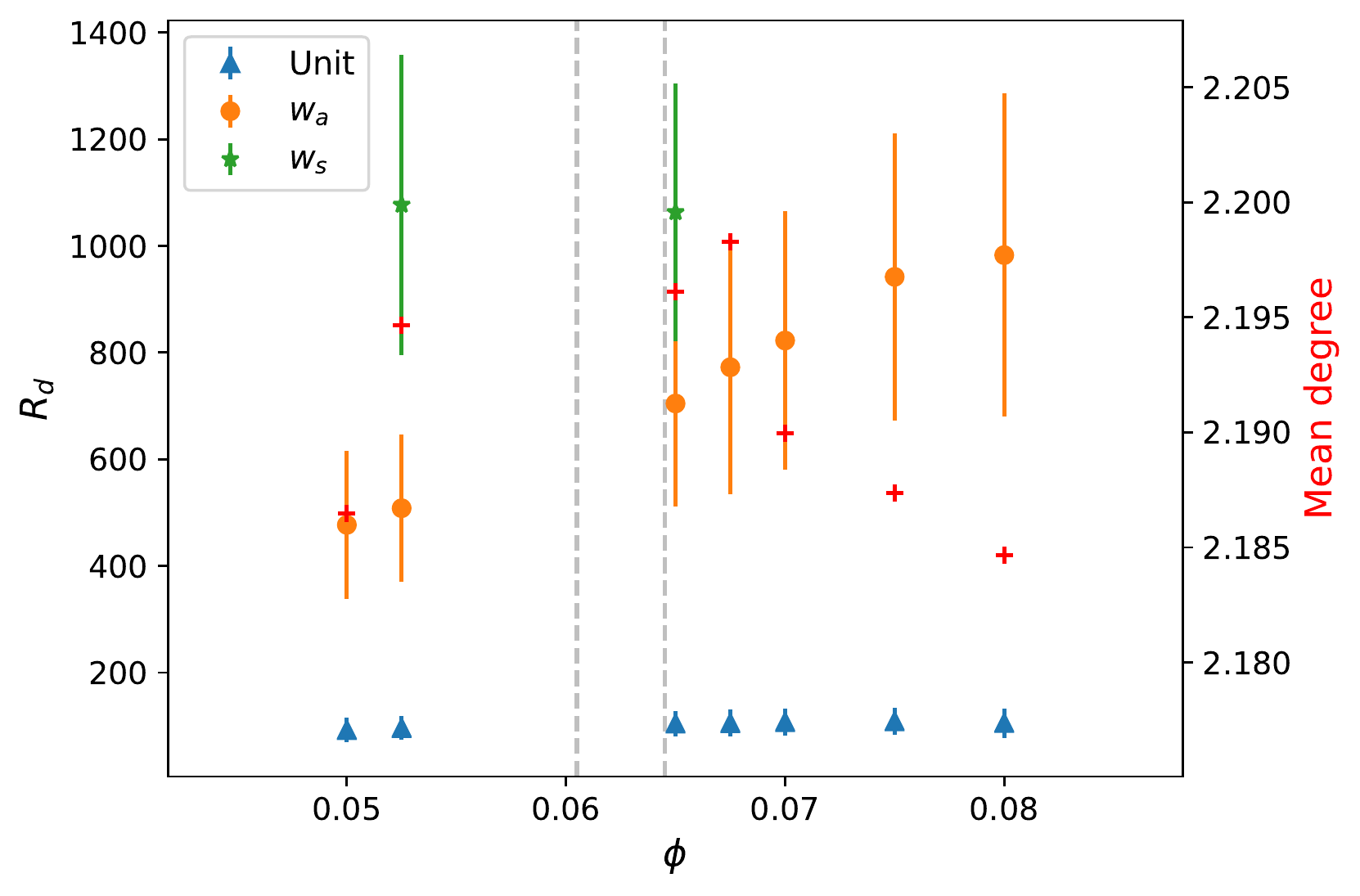}
\caption{Comparing the dependence of the effective resistance between furthest nodes of the network $R_d$ on $\phi,$ and for $D/L=50,$ according to: unit conductances (labelled \textit{Unit}, triangle, blue), the anisotropic model $w_a$ Eq.~\ref{eq:geomstack} (circle, orange) and the surface-surface overlap model $w_s$ (star, green). The mean node degree for the sampled largest clusters of each $\phi$ is shown on the second Y-axis (plus, red). The dashed vertical lines indicate the simulation estimates of the I-N coexistence region.}
\label{fig:Rdiam50}
\end{figure}

\begin{figure}
\centering
\includegraphics[width=1.0\linewidth]{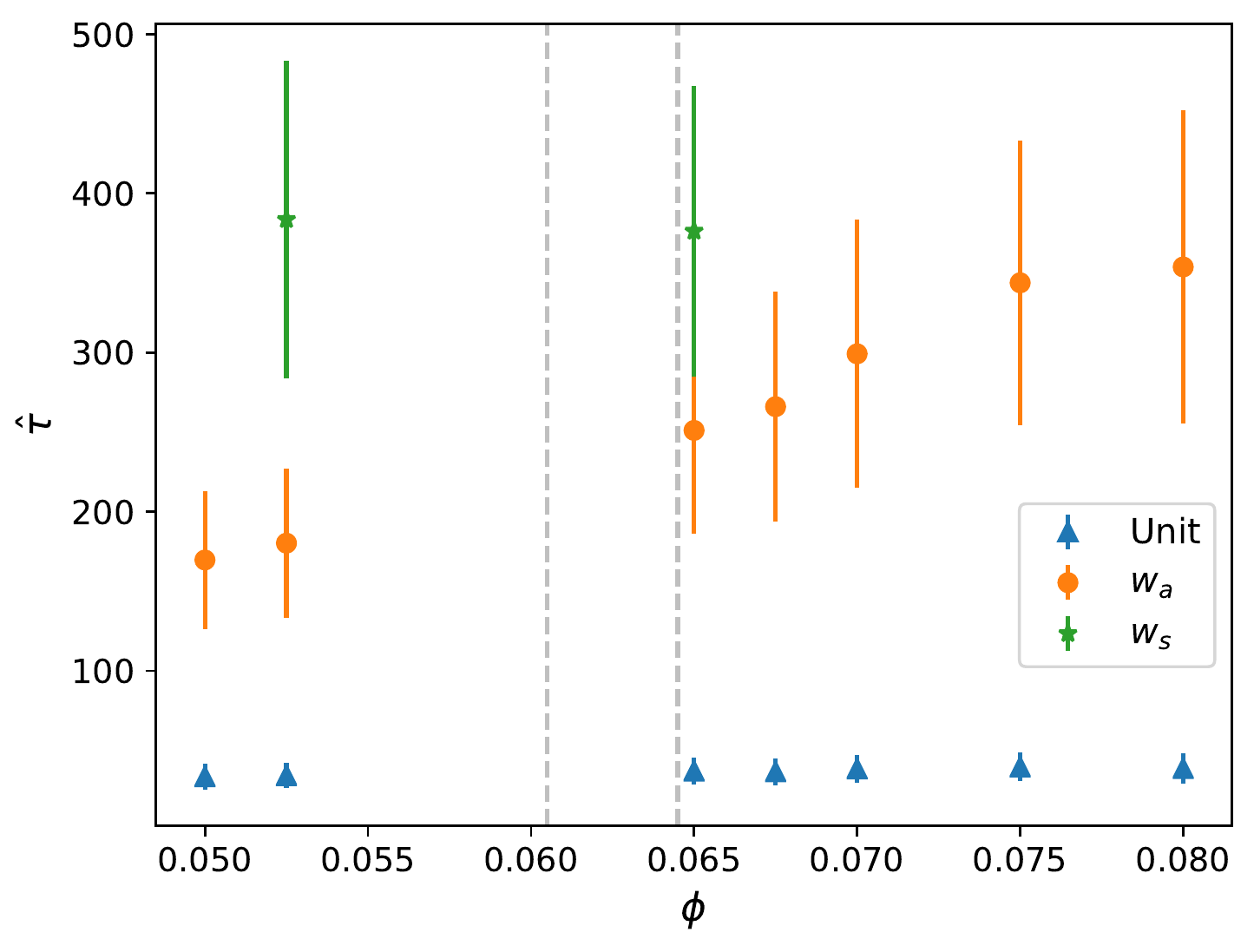}
\caption{Normalized Kirchhoff index of the tunneling networks (for $D/L=50$) as a function of volume fraction for unit conductances (labelled \textit{Unit}, triangle, blue), and for the anisotropic model $w_a$ Eq.~\ref{eq:geomstack} (circle, orange).}
\label{fig:taudegree50}
\end{figure}

\clearpage

% % Create the reference section using BibTeX:
%\bibliography{Bibliography}

\end{document}